\documentstyle[prl,aps]{revtex}
\begin{document}
\title
{Non-occurrence of Trapped Surfaces and  Black Holes in
Spherical Gravitational Collapse}

\author{Abhas Mitra}
\address{Theoretical Physics Division, Bhabha Atomic Research Center,\\
Mumbai-400085, India\\ E-mail: amitra@apsara.barc.ernet.in}


\maketitle

\begin{abstract}
We carefully analyze the apparently commonplace yet  subtle concepts
associated with the notion of existence of Black Holes.
We point out that although the pioneering work of
Oppenheimer and Snyder (OS),
technically,
indicated the formation of an event horizon for a collapsing homogeneous dust
ball of mass $M_b$ in that the
 circumference radius the outermost surface, $R_b\rightarrow
R_{gb}= 2GM_b/ R_b c^2$, in a proper time $\tau_{gb} \propto R_{gb}^{-1/2}$,
it never explicitly showed the
formation of a ``trapped surface'' where $R_b <R_{gb}$. On the other hand, the
Eq. (36) of their paper ($T\sim \ln{y_b+1\over y_b-1}$)
 categorically demands that $y_b=R_b/ R_{gb} \ge 1$, or $2G M_b/R_b c^2
\le 1$, so that, if
one has to pursue the central singularity, $R_b\rightarrow 0$, it is
necessary that $R_{gb} \rightarrow 0$ at a faster rate. Consequently,
actually, $\tau_{gb} \rightarrow \infty$, and more importantly, the
(fixed) gravitational mass of the dust $M_b=0$. Further, by analyzing the
general inhomogeneous dust solutions of Tolman
in a proper physical perspective, we
show that, all dust solutions obey the same general constraint and
are characterized by $M_b=0$. Next, for a
collapsing fluid endowed with  radiation pressure, in a {\em most general
fashion}, we
 discover that the collapse equations obey the same Global Constraint $2GM
/R c^2 \le 1$ and which specifically shows that, contrary to the
traditional intuitive Newtonian idea, which equates
the gravitational mass ($M_b$) with the fixed baryonic mass ($M_0$),
 the {\em trapped
surfaces} are not allowed in general theory of relativity (GTR). Now by
invoking, the ``positive mass theorems'', it follows that for continued collapse,
the final gravitational mss $M_f \rightarrow 0$ as $R\rightarrow 0$.
Thus we confirm Einstein's and Rosen's idea that  Event Horizons and
Schwarzschild Singularities are unphysical and can not occur in Nature.
This,
in turn, implies that, {\em if there would be any continued collapse}, the
 initial gravitational mass energy of the fluid must be radiated away
$Q\rightarrow M_i c^2$. \footnote{gr-qc/9810038}

{\em Irrespective of the gravitational collapse problem}, by analyzing the
properies of the Lemaitre and Kruskal transformations, in a
straightforward manner, we show that finite
mass Schwarzschild Black Holes can not exist at all.
\end{abstract}

\newpage

\section{Introduction}
One of the oldest and most fundamental problem of physics and astrophysics
is that of gravitational collapse, and, specifically, that of the ultimate
fate of a sufficiently massive collapsing body. Most of the astrophysical
objects that we know of, viz. galaxies, stars, White Dwarfs (WD), Neutron
Stars (NS),  in a broad sense, result from gravitational collapse. And
in the context of classical General Theory of Relativity (GTR), it is believed
that the ultimate fate of sufficiently massive bodies is collapse to a
Black Hole (BH). A spherical chargeless BH of (gravitational) mass $M_b$
is supposed to occupy a region of spacetime which is separated by a
hypothetical one-way membrane of ``radius'' $R_{gb} = 2G M_b/c^2$, where $G$
is the Newtonian gravitational constant and $c$ is the speed of light. This
 membrane, called, an event horizon, is supposed to contain a central
singularity at $R=0$, where most of the physically
relevant  quantities like (local) energy density,
(local) acceleration due to gravity, (local) tidal acceleration,
and components of the Rimmenian curvature tensor diverge.
However, although such ideas are, now, commonly believed to be elements of
ultimate truth, the fact remains that, so far, it has not been possible to
obtain any
analytical solution of GTR collapse equations for a physical fluid endowed
with pressure ($p$),  temperature $(T)$ and an equation of state (EOS).
And the only situation when these equations have been solved (almost) exactly,
is by setting $p\equiv0$, and further by neglecting any density
gradient, i.e., by considering
$\rho = constant$ [26]. It is
believed that these
(exact) asymptotic solutions actually showed the formation of BH in a
finite comoving proper time $\tau_{gb}$. However, this,
assumption of perfect homogeneity is a very special case, and, now many
authors believe that for a more realistic inhomogeneous dust, the results
of collapse may be qualitatively different. These authors, on the strength
of their semi-analytical and numerical computations, claim that the
resultant singularity could be a ``naked'' one i.e., one for which there
is no ``event horizon'' [7, 12, 13, 22 36].
 Therefore light may emanate from a naked
singularity and reach a distant observer. A naked singularity may also
spew out matter apart from light much like the White Holes.
In other words, unlike  BHs, the
naked singularities are visible to a distant observer and, if they exit, are
of potential astrophysical importance. However, according to a celebrated
postulate by Penrose [42], called ``Cosmic Censorship Conjecture'', for
all realistic gravitational collapse, the resultant singularity must be
covered by an event horizon, i.e, it must be a BH. And many authors
believe that the instances of occurrences of ``naked singularities'' are
due to fine tuned artificial choice of initial conditions or because of
inappropriate handling and interpretation of the semi-analytical
treatments. In this paper, we are not interested in such issues
 and would avoid
presenting and details about the variants of naked singularities (strong,
weak, local, global, etc) or the variants of the censorship conjecture. Also in this
paper, we are  interested only in the case of collapse of physical matter consisting
of baryons and leptons, and would completely avoid any discussion on
collapse of hypothetical fields like various ``scalar fields''.

When we say that BHs result from collapse of ``sufficiently'' massive
bodies, it is in order to qualify the term ``sufficient''. Very crudely,
the cores of moderately massive stars end up as WDs, a configuration
supported by pressure of degenerate electrons. And it was shown by Chandrasekhar
[46] and independently by Landau [29]
that there is an upper limit on the mass of
the WDs, called, ``Chandrasekhar Mass''
\begin{equation}
M_{ch} = {3.1 \over \mu_e^2} \left({\hbar c\over 2\pi G}\right)^{3/2}
\approx 1.457  \left({2\over \mu_e}\right)^2 M_{\odot}
\end{equation}
where $\mu_e$ is the number of electrons per nucleon, so that, for a He-
WD, $\mu_e =2$
When the main sequence mass of a star is such that, the final mass of its core
$M_c > M_{ch}$, the core continues to collapse without ever resting in a state
of hydrostatic equilibrium supported by degenerate electron pressure. At a
(baryon) density of $\sim 10^{11}$ g/cm$^3$, neutronization of matter
starts, and a new state of hydrostatic equilibrium may be reached where the
pressure is due to degenerate neutrons. In other words, the collapse
process ends with the formation of a NS. But again, there is an upper
limit on the mass of stable NSs, called, Oppenheimer and Volkoff limit,
$M_{OV}$. The original value of this limiting mass was  first obtained by
Oppenheimer and Volkoff (1938) by treating the NS as a self-gravitating gas
of free neutrons, $M_{OV} \approx 0.7 M_{\odot}$. However, in the past few
decades, with the progress of nuclear physics, there have been enormous
amount of work to find the value of $M_{OV}$ using actual EOS of nuclear matter.
We would only mention here a particular value obtained by using an EOS
which incorporates the fact the sound speed in nuclear matter is limited
by the speed of light, $dp/d\rho \leq c^2$, [8, 53]
\begin{equation}
M_{OV} \approx 3.2 M_\odot \left({5 \times 10^{14}
 {\rm g cm}^{-3} \over \rho}\right)^{1/2}
\end{equation}
Thus, more massive stellar cores are supposed to undergo ``continued collapse''
without reaching a new state of hydrostatic equilibrium, and are believed
to end up as BHs (or naked singularities). However, to fully appreciate
this conviction it is necessary to understand another point. All along
this chain of previous discussion it was implicitly assumed that, during
the collapse process, the role of GTR is negligible except for the stage
beyond the NS stage, and the instantaneous gravitational mass of the core
\begin{equation}
M_f \approx M_i \approx M_b= M_0 = N m
\end{equation}
where the constant the baryonic mass of the core (by ignoring the mass of the
leptons and assuming no anti baryons to be present) $M_0 = m N$, with $m$
as the mean nucleon mass and $N$ to be the number of nucleons. Under this assumption,
an event horizon is formed at a density
\begin{equation}
\rho = {3c^6\over 32 \pi G^3 M^2} \approx 2\times  10^{16} {\rm g~cm}^{-3}~ \left({M_0\over
M_{\odot}}\right)^{-2}
\end{equation}
So once the collapse proceeds beyond nuclear densities, $\rho_{nu} \sim
10^{15}$ g cm$^{-3}$, very soon, it is expected that a surface will be
formed where $2 GM/R c^2 >1$ from within which even light cannot escape
[41]. Since radiation can not escape  the gravitational mass
of the core remains fixed at $M_g = M$. For further appreciation of this
introductory note, it is necessary to have a clear idea about the notion
about the evolution of the Gravitational Binding Energy (``mass defect'')
during gravitational contraction.
\subsection {Kelvin - Helmholtz (KH) Process}

In GTR, all the global concepts associated with the notion of energy,
like the total gravitational mass of a (spherical) body, can be
meaningfully defined only with respect to a distant observer $S_\infty$ who
does not feel any gravitation. Further, the total energy of the body  is the sum total of all kinds of associated energy
(as
measured by $S_\infty$):
\begin{equation}
E \equiv M_b c^2 = M_0 c^2 + E_N + E_{kinetic}
\end{equation}
where $E_N$ is the total energy in the Newtonian sense, i.e., excluding,
the rest mass contribution $M_0 = m N$ and $E_{kinetic}$ is the energy
associated with accelerated bulk motion.
Also, for a spherical body,
where the mass (energy) contained within a radius $R$, and is defined with
respect to
$S_\infty$ is
\begin{equation}
M(R) = \int_0^R \rho dV =\int_0^R (\Gamma_s \rho) d{\cal V}
\end{equation}
so that
\begin{equation}
M_b = \int_0^{R_b} \rho dV= \int_0^{R_b} (\Gamma_s \rho) d{\cal V}
\end{equation}
It should be remembered here that $dV =4\pi R^2 dR$ is not the
the physical volume element (measured locally), and the latter is given by
\begin{equation}
d{\cal V} = {dV\over \Gamma_s}
\end{equation}
where
\begin{equation}
\Gamma_s (R) = \left( 1- {2 G M(R) \over R c^2}\right)^{1/2}
\end{equation}
It may be verified that in the limit of weak gravity, i.e., $2GM/Rc^2 \ll
1$, $d{\cal V} \rightarrow dV$.
Here the  bulk kinetic energy term is actually inseparably connected
with the definition of $M_b$, and can be separated only if
gravity is not too strong. And we shall see later how to  define $M_b$
by taking into account dynamic motion in an organic fashion.

As energy is released by the collapse process,
the body becomes gravitationally more tight, and its gravitational binding
energy is defined as
\begin{equation}
-E_B= B.E =  M_f - M_0
\end{equation}
where $M_0= N m$ is the gravitational energy when the body may be imagined
to be dispersed to infinity. If we add all the energy liberated since
this state of infinite dilution, then $M_i = M_0$ and
\begin{equation}
-B.E. = (M_f -M_i)c^2 = Q
\end{equation}
For formation of a NS in a typical
type II supernova event, it is found that, the amount of energy released
in the form of neutrinos $Q \sim 3 \times 10^{53}$ ergs. And thus,
actually, the gravitational mass of the NS is smaller than the core
preceding it by an amount $Q/c^2 \sim 0.1 M_\odot$.
Sometimes, it is loosely mentioned that, the B.E. is the gravitational
potential energy or self-gravitational energy of the body, $E_{ g}$. This is
erroneous, although, by virtue of the Virial Theorem (VT), it may turn out
that, the $B.E. \sim  E_{ g}$. The B.E. is actually, the total energy
of the body defined in the Newtonian sense, i.e., by excluding any rest mass
\begin{equation}
-E_B \equiv E_N = E_{ g} + E_{ in} = (M_f -M_0) c^2
\end{equation}
where, $E_{in}$ is the internal or thermodynamical energy of the body.
In Newtonian gravity, the
self-gravitational energy of a polytrope of degree $n$ is:
\begin{equation}
E_{ g}^N = -{3\over 5-n} {G M_b^2 \over R_b}
\end{equation}
For a homogeneous sphere, $n=0$, and we find
\begin{equation}
E_{ g}^N = -{3\over 5} {G M_b^2 \over R_b}
\end{equation}
The Newtonian form of Virial Theorem states that, if $\gamma_t$
(thermodynamical) is the
effective ratio of specific heats of the self-gravitating gas, it is found
that [47]
\begin{equation}
E_g + 3(\gamma_t -1) E_{in}=0
\end{equation}
Or,
\begin{equation}
E_{in} = {-E_g \over 3(\gamma_t -1)}
\end{equation}
By combining  Eqs. (1.12, 1.15 and 1.16) we find
\begin{equation}
E_N= {3\gamma_t -4\over 3(\gamma_t -1)} E_g
\end{equation}
Both for  a monoatomic non-degenerate perfect gas and a cold degenerate gas,
we have $\gamma_t =5/3$, so that
\begin{equation}
E_N = {1\over 2} E_g = - {1\over 2} \mid E_g\mid
\end{equation}
As the system contracts, the amount of $pdV$ work is performed by self-gravity
is $\mid \Delta E_g\mid$. Out of this, only a certain fraction is utilized
in increasing the internal energy:
\begin{equation}
\Delta E_{in} = {\mid \Delta E_g\mid \over 3(\gamma_t -1)}
\end{equation}
And the rest of the energy must be radiated away:
\begin{equation}
Q = - \Delta E_N = {3\gamma_t -4\over 3(\gamma_t -1)} \mid \Delta E_g\mid
\end{equation}
Thus {\em emission of energy from a contracting self-gravitating body is a
necessary and inescapable phenomenon}. As the body contracts, it emits
energy and yet  tends to be hotter because of the increase of
 internal energy - this property is known as the ``negative'' specific
heat of gravity. As a result, in the absence of other sources of energy
(like nuclear fusion energy), gravitational contraction tends to be a runaway
process and can be absolutely halted only if $\gamma_t \rightarrow
4/3$. For a  system comprising substructures like atoms and nuclei, the
contraction process would release new degrees of freedom and the value of
$\gamma_t \rightarrow 4/3$ or even,  momentarily, be $<4/3$ even when all
the constituent particles are not in a state of extreme relativistic
degeneracy. After the neutronization process, such a thing happens during
the supernova collapse prior to the attainment of nuclear density of the
collapsing matter; and the collapse during this stage is near adiabatic
[47]. But in the limit of a monoatomic gas,
when new degrees of freedom are not suddenly liberated,
 the value of $\gamma_t \rightarrow 4/3$ only if
 all the fluid particles become relativistically degenerate with individual
momenta  $\rightarrow \infty$. Thus, this can happen only
asymptotically, and, very strictly, it can not be exactly realized except at a
physical singularity.
It may appear that, if the fluid is buried under an event
horizon and yet $R$ is finite, the emission of radiation will stop because then the fluid can
not communicate with $S_\infty$. However, by the Principle of Equivalence,
to be elaborated latter, the local laws of thermodynamics remains
unchanged, and although, it is not possible to define $E_g$ meaningfully
in such cases, it is possible that the fluid will still require to radiate
to honor thermodynamics. This difficulty can be alleviated if we assume
$\gamma_t =4/3$ inside the event horizon. But, the pressure, internal energy
and all other physical quantities remain finite for a finite $R$, and, as
discussed above, $\gamma_t$ should actually  be $> 4/3$.

The virial theorem (VT) used in the present discussion
 is actually  due to the Newtonian
inverse square law of gravitational acceleration. There is no counterpart
of a GTR VT, in an easily usable form, yet. It is interesting to
recall that, for spherical symmetry and
in the absence of any  angular momentum of the ``test particle'', the
effective potential felt by a test particle still has this Newtonian
$R^{-1}$ form. In
general, all the peculiarities associated with gravity get accentuated by
GTR, and it is  possible that a VT almost similar to the one used here
might be
applicable in the GTR case too.
However, when GTR becomes important, this Newtonian expression for
gravitational energy has to be modified. In GTR, like all global energies,
$E_g$ too is defined only with respect to $S_\infty$
[9, 48] :
\begin{equation}
E_g = \int_0^{R_b} {\Gamma_s -1\over \Gamma_s} \rho  dV =
\int_0^{R_b} (\Gamma_s -1) \rho  d\cal{V}
\end{equation}
It may be verified that in the limit of weak gravity, i.e., $2GM/Rc^2 \ll
1$, the GTR expression for $E_g$ is reduced to the Newtonian form.

The internal energy can have two contributions:
\begin{equation}
E_{in} = E_{T} + E_{cold}
\end{equation}
where $E_{T}=\int e_{T} d\cal{V}$ is the temperature
dependent thermal part of the
internal energy and $E_{cold}= \int e_{cold} d\cal{V}$ is due to the
 pure degeneracy effects and which may
exist in certain cases even if the star is assumed to be at a temperature
$T =0$. The corresponding energy densities are
\begin{equation}
e_T = {(3/\pi^2)^{1/3} mc^2\over 6 (\hbar c)^2} n^{1/3} T^2; \qquad
T\rightarrow 0
\end{equation}
Actually, when the body is really degenerate, this kind of
splitting of $E_{in}$ can be done only in an approximate manner. For
example, if it is assumed that a degenerate ideal neutron gas is close to
$T=0$, i.e, $ T \ll T_{Fermi}$, then one may approximately take the first
term  (lowest order in $T$) of an infinite series to write the above expression.
On the other hand, if the temperature is indeed much higher than the
corresponding Fermi temperature, degeneracy will vanish, $e_{cold}
\rightarrow 0$, and the entire energy density will be given by the thermal
contribution:
\begin{equation}
e= e_{T} = {3\over 2} n kT; \qquad T\rightarrow \infty
\end{equation}
and
\begin{equation}
e_{cold}= {2\over 3} p_{cold} = {2 (3\pi^2)^{2/3}\over 15} {\hbar^2\over
m} n^{5/3}; \qquad \gamma_t =5/3
\end{equation}
Since it is not known beforehand, how $T$ would evolve, in principle, one
should work with an expression for $e_T$ (an infinite series) valid for
arbitrary $T$. But, it  is not possible to do so even for an ideal Fermi gas. As to the
actual EOS of nuclear matter at a finite $T$, it may be remembered that,
it is an active field of research and still at its infancy. Thus, in practice it is
impossible to make much headway without making a number of simplifying
assumptions because of our inability to {\em self consistently} handle:
(i) the equation of state (EOS) of matter at arbitrarily high density and
temperature, (ii) the opacity of nuclear matter at such likely unknown
extreme conditions, (iii) the associated radiation transfer problem and
all other highly nonlinear and coupled partial differential (GTR)
equations (see later).

One may start the numerical computation by presuming that indeed the
energy liberated in the process $Q \ll M_i c^2$, i.e., the effect of GTR
is at best modest. Then, it would naturally be found that the temperature
rise is moderate and depending on the finite grid sizes used in the
analysis and limitation of the computing machine,
 one may conclude that the formalism adopted is
really satisfying, and then find that $Q \ll M_i
c^2$ [14, 51, 52, 54].
 Meanwhile, one has to extend the  presently known (cold) nuclear EOS at
much
higher densities and maintain the assumption that the rise
in temperature is moderate. Because if $T$ is indeed high, in the
diffusion limit, the emitted energy $Q \sim T^4$ would be very high, and
the value of $M_f$ could drop to an alarmingly low value. Thus,
 for the external spacetime, one needs to consider the Vaidya
metric [34]. Actually, even
when, $T$ is low, it is extremely difficult to self consistently handle
the coupled energy transport problem.

It may appear that, the practical difficulties associated with the study of
 collapse involving densities much higher than the nuclear density can be
avoided if one starts with a very high value of $M_i$, say, $10^{10}
M_\odot$. Then {\em if one retains the assumption} that $M_i = M_f$, one would
conclude that an ``event horizon'' is formed at a density of $\sim 10^{-4}$ g
cm$^{-3}$, where the EOS of matter is perfectly well known. {\em What is
overlooked} in this traditional argument is that, once we are assuming that
an event horizon is about to form, we are endorsing the fact that we are
in the regime of extremely strong gravity, and, therefore for all the
quantities involved in the problem,  a real GTR
estimate has to be made without making any prior Newtonian approximation.
To further appreciate this important but conveniently overlooked point by
the numerical relativists note that the strength of the gravity may be
approximately indicated by the ``surface redshift'', $z_s$, of the collapsing
object, and while a Supermassive Star may have an initial value of $z_s$
as small as $10^{-10}$, a canonical NS has $z_s \sim 0.1$, while the Event
Horizon, irrespective of the initial conditions of the collapse, has got
$z_s =\infty$! Therefore all Newtonian or Post Newtonian estimates or the
conclusions based on such estimates have {\em little relevance for actual
gravitational collapse problem}.

We
would see in the latter part of this paper that, for a ``test particle''
in the External Schwarzschild metric [28], the
maximum value of the local free fall speed of the fluid appears to be
$v_{ex} = (R_g/R)^{1/2} c$. And $v_{ex} \rightarrow c$ as
$R\rightarrow R_g= 2GM/  c^2$.
 But then for a finite value of $R_g = 2GM/c^2$, it is
{\em believed that this anomalous behavior results from} a ``coordinate
singularity''. When, this ``coordinate singularity'' is removed, it is
{\em expected} that {\em the
actual value of} $v$ would $ \rightarrow c$ {\em only when the fluid collapses to the
central singularity} $R \rightarrow 0$. Therefore, the actual value of $v$
must be considerably less than light speed at $R=R_g$, if $R_g \neq 0$.
For a fluid endowed with pressure, the collapse process is bound to be
slower, and therefore, one may legitimately expect $v $ to be appreciably lower than
$c$ for $ R \ge R_g$, and consequently, the bulk flow kinetic energy to be
considerably smaller than $M c^2$ (if $R_g >0$). Then one  may
crudely use the Eqs. (1.9) and (1.21) to find that
 $\Gamma_s \sim 0$, so that $\mid E_g\mid $
{\bf increases drastically}. As a result, the integrated value of $Q$ would tend
to increase drastically, and this would {\em pull down the running value of}
$M_f = M_0 - Q/c^2$ and $R_{gb}$ to an alarming level!
In fact Eq. (1.6) indeed shows that the value of $M$ should drop
significantly as $\Gamma_s$ is excepted to decrease substantially near the horizon
irrespective of the value of $M_0$ and $M$. At the same time, of course,
 the
value of $R$ is decreasing. But how would the value of $R_b/R_{gb}$ would
evolve in this limit? Unfortunately, nobody has ever, atleast in the
published literature, tried to look at the problem in the way it has been
unfolded
above. On the other hand, in Newtonian notion, the value of $M_f$ is
permanently pegged at $M_0$ because energy has no mass-equivalence
(although in the corpuscular theory of light this is not so, but then
nobody dragged the physics to the $R\rightarrow R_g$ limit seriously then).
So, in Newtonian physics, or in the intuitive thinking process of even the
GTR experts, the value of
\begin{equation}
{2 G M \over R c^2} \equiv {2G M_0 \over R c^2} \rightarrow \infty;
\qquad R\rightarrow 0
\end{equation}
 and the idea of a trapped surface seems to be most natural.
But, in GTR, we can not say so with absolute confidence even if we start
with an arbitrary high value of $M_i$ because, in the immediate
vicinity of $R \rightarrow R_{g}$, the running value of $M$ may
decrease in a fashion which we are not able to fathom either by our crude
 qualitative arguments, based on GTR, or by numerical computations
plagued with uncertain physics and inevitable machine limitations.

And, note, the Eq.(1.2) should actually be modified to:
\begin{equation}
\rho_g = {3c^6\over 32 \pi G^3 M^2} \approx 2 \times 10^{16} {\rm g~cm}^{-3}~ \left({M_f\over
M_{\odot}}\right)^{-2}
\end{equation}
And, if $M_f$ drops to an alarming level, the actual value of $\rho_g$
can rise to very high values. Thus all the
 difficulties associated with the numerical study of  the collapse of a stellar
mass object
 may reappear  for any value of $M_0$  unless one hides the
nuances of GTR and other
detailed physics with favorable and simplifying assumptions and approximations.

Our treatment of GTR in this introductory section has been crude and
inaccurate. In fact, had we pursued the evaluation of $E_g$ in the
$R\rightarrow R_g$ limit in this crude manner, we would have found
$M_f$ to be negative (we shall see later that it is a valid quest in GTR
to see if $M$ can be negative). To seek a real answer for such questions,
we need to handle GTR carefully and exactly in a manner different from
this qualitative approach. Before we proceed to do that, it would be
worthwhile to briefly recall the gravitational collapse problem in the
context of the Newtonian physics.
\section{ Newtonian Spherical Collapse}
As already emphasized, in Newtonian gravity
\begin{equation}
M_i=M_f=M_0 = m N= constant
\end{equation}
where $N$ is the total (fixed) number of baryons, also,
 $\rho \equiv \rho_0 =m n$, where $n$ is the baryon number density.

Thus, long time ago, it was envisaged by Michell in 1784 [24] and independently
by Laplace [37] that, as a massive gas cloud contracts, at a certain stage,
one has $R_b=R_{gb}$, when the escape velocity would become equal to  $c$,
 and then
even light would not be able to escape the collapsing body:
\begin{equation}
v_{escape} = \left({2 G M_b\over R_{gb}}\right)^{1/2} = c
\end{equation}
After this, we would find $v>c$ and $\rightarrow \infty$.
The fluid would collapse to a singularity $R=0$ in a finite time $t_c< R/c$.
Actually, however, the value of $t_c$ can not be evaluated exactly
in Newtonian physics unless one considers the gas cloud to be a
homogeneous dust! The central equation for Newtonian collapse is the Euler
equation:
\begin{equation}
\rho {dv \over dt} = - \rho {G M(R) \over R^2} - {dp \over dR}
\end{equation}
Since the EOS of the fluid can never be known at an arbitrary density and temperature,
there is no question of exactly solving the collapse problem either in GTR
or even in Newtonian case. Although, in the Newtonian case $M$ does not
decrease because of emission of radiation, one should, strictly, add the
term due to radiation pressure gradient $dp_r /dR$ in the foregoing
equation to inadvertently make the case even more intractable. The only
way one may hope to solve this equation would be to assume $p\equiv 0$,
i.e., to consider the fluid to be a dust. One may try to justify this
assumption in the following way:

Assume $T(R, t) = T_0= constant$, i.e., ignore the KH process. Then the
EOS simplifies to
\begin{equation}
p = {k \rho T_0 \over m} \propto \rho
\end{equation}
where $k$ is the Boltzman's constant.  Further approximate the last term of
the  Eq. (2.3) as $dp/dR \sim p/R$. But the first term on the right hand side of the
same Eq. is $\propto R^{-2}$, because $M(R)$ is constant. Then as $R$ decreases,
at a certain stage, the pressure gradient term may become insignificant
compared to the gravitational acceleration term. And then one might treat
the fluid as a ``dust''. Yet for an exact solution, it is necessary to
assume the dust to be homogeneous unless we impose the condition of self-similarity.
For a homogeneous dust, we have
\begin{equation}
M(R) = {4 \pi R_\infty^3 \over 3} \rho(0)= {4 \pi R^3 \over 3} \rho
\end{equation}
where $\rho(0)$ is the density at an initial radius $R_\infty$ for a given layer.
Then the collapse equation reduces to
\begin{equation}
{\ddot R} = -{K \over R^2}
\end{equation}
where
\begin{equation}
K = {8\pi R_\infty^3 G \rho(0) \over 3}
\end{equation}
By multiplying both sides of the above equation with $v={\dot R}$ and
integrating, we obtain
\begin{equation}
v^2 = 2K \left( {R_\infty\over R} -1\right)
\end{equation}
where it has been assumed that {\em the dust was initially at rest}:
\begin{equation}
v(R, 0) =0; \qquad R= R_\infty
\end{equation}
Without this foregoing assumption, the problem remains imprecisely
defined. However, by using the $p=0$ EOS in Eq. (2.3) , it can be found that, a
dust can never be at rest, and, technically, for a correct description, we
should take $R_\infty = \infty$. In that case, the problem would be that, the
dust solution can not be smoothly matched onto the collapse solution for a
star of finite radius. This is a fundamental inconsistency of all dust
solutions and it only asserts the fact that a physical fluid can never,
really, behave as a dust. We ignore this difficulty and set
\begin{equation}
\chi \equiv {R\over R_\infty}= \cos^2 \beta
\end{equation}
in Eq. (2.8) to obtain
\begin{equation}
K^{1/2} t = \beta + {1\over 2} \sin 2\beta
\end{equation}
At the beginning of the collapse, we have $\chi =1$ and $\beta =0$, and
the collapse terminates at $\chi =0$ and $\beta =\pi/2$. Therefore the
value of $t_c$ is
\begin{equation}
t_c = {\pi\over 2\sqrt {K}}= {\pi \over 2} \left( {3\over 8 \pi G \rho(0)}\right)^{1/2}
\end{equation}
We shall see later that, in GTR too, one obtains {\em exactly the same
expression} for $t_c$ !
If this equation is to taken seriously, a star like Sun with a present
density of $\rho \sim 1$ g cm$^{-3}$, would collapse within a time of $t_c
\sim 30$ minutes. Obviously such a thing does not happen. But is it
because of the fact that Sun burns nuclear fuel at its center to maintain
a pressure gradient? What, if the source of nuclear fuel is suddenly
removed or exhausted? Could Sun behave like a dust,  and collapse within
30 minutes? If we do not invoke any probable degeneracy pressure,
certainly, Sun would start contracting, but this does not at all mean
that, it would undergo free fall. The KH process would
always maintain a pressure gradient as long as $\gamma_t > 4/3$ (as long
as we treat as a simple mono atomic gas). And because of the same process
Sun would continue to radiate even in the absence of any nuclear fuel.
The approximate value of the gravitational energy of Sun is
\begin{equation}
\mid E_{g\odot}\mid \sim {G M_{\odot}^2 \over R_{\odot}}\sim  4 \times 10^{48} {\rm erg}
\end{equation}
And by VT, the total Newtonian energy or the B.E. of Sun is
\begin{equation}
E_\odot = {1\over 2} E_{g\odot}
\end{equation}
If the Sun continues to radiate at its present rate of $L_\odot \approx 4
\times 10^{33}$ erg, a very crude estimate of the time of collapse would be
the so-called KH- time scale [46].
\begin{equation}
t_\odot \sim {\mid E_\odot\mid \over L_\odot} \sim 5 \times 10^{14} {\rm
s} \sim 10^7 {\rm yr}
\end{equation}
Here, it may be argued that, since, the Kelvin- Helhmoltz process is a
runaway process, the value of $L$ would increase rapidly and the time of
collapse would decrease. This is not necessarily so because, as $R$
decreases, the value of $\mid E_g\mid$ constantly increase till we invoke
the non - Newtonian idea that as a result of radiation, the value of $M$
keeps on decreasing too. It should be clear that we are really not
interested here in finding the precise value of $t_c$ for any problem by
using imprecise Newtonian physics, and, on the other hand, all we wanted to show is that
{\em the assumption of dust collapse may present a completely erroneous picture
for a
 real physical problem unless the initial conditions are set self-consistently}.
 As mentioned before, the Eq. (2.3)  strictly demands either
 \begin{equation}
 R_\infty =\infty ; \qquad \rho =finite; \qquad M=\infty
 \end{equation}
 Or else
 \begin{equation}
 R_\infty = finite; \qquad \rho =0; \qquad M =0
 \end{equation}
 In fact in  both the cases, we would find $t_c =\infty$. We
  do not claim here that it is actually so, but, we are only emphasizing, what
the value of $t_c$ really should be in the dust  model if we put the
initial conditions in a physically valid
self-consistent manner. It is also interesting to go back to the supposed
turning point when the gravitational acceleration overtakes the pressure
gradient term. It may be found that for a stellar mass object
this would happen at
\begin{equation}
kT \sim {G M m \over R} \sim 10 {\rm GeV} \left({R \over 10^6 {\rm cm}}\right)^{-1}
\end{equation}
If at this stage $\gamma_t \rightarrow 4/3$,  and the system starts to
evolve adiabatically (all dust solutions are necessarily adiabatic), and
 the temperature would be
rising as $T \sim R^{-1}$. Of course, the perfect gas EOS would break down
long before  such a thing would happen.
\section { Elements of GTR}
Both the Special Theory of Relativity (STR) and General Theory of Relativity
(GTR) considers the spacetime as a  differentiable 4-dimensional (pseudo)
Rimmenian manifold
 with each point of the manifold corresponding to an event in
spacetime. Both in STR and GTR, a metric is the {\em invariant} distance
between two nearby events. In STR, we have
\begin{equation}
ds^2 = \eta_{ik} dx^i dx^k
\end{equation}
where Latin indices run from (0, 3) and Greek ones run from (1, 3). Here,
\begin{equation}
\eta_{ik} = {\rm diag} (1, -1, -1, -1)
\end{equation}
has a signature of $-2$. One can also work with a signature of $\eta_{i
k}$ as $+2$. The spacetime of STR is ``flat'' because $\eta_{ik}$
is independent of $x^i$. In contrast, in GTR, the metric coefficients are
coordinate dependent, and this coordinate dependence can not be (globally)
eliminated by any coordinate transformation; and hence
the spacetime is ``curved'':
\begin{equation}
ds^2 =g_{ik} (x^i) dx^i dx^k
\end{equation}
Yet, by the Principle of Equivalence (POE) any infinitesimally small patch
of spacetime can always be considered locally flat, i.e., POE asserts
that, in the infinitesimal neighborhood of any event there exists a
local Inertial (Lorentz)  Frame (LIF) with respect to which
\begin{equation}
g_{ik} \rightarrow \eta_{ik}; ~(\rm locally)
\end{equation}
and all Christoffel symbols vanish (locally).
 Without the
existence of POE and LIF, GTR has hardly got any practical utility. A
stronger version of POE asserts that all non-gravitational laws of physics
studied in STR remains (locally) valid in LIF.
 While the previous version of POE
implicitly assumes that {\em it is possible to define a local {\em speed} $v$ of the
particle or fluid under consideration} whose value is less than $c$, the
{\em stronger version of the same specifically demands so}.

 One of the foundational aspects of  STR is that unlike an Euclidean space,
the  metric of relativity $ds^2$ is not positive definite and the worldlines,
i.e., trajectories of events in the spacetime
 can be classified as
 \begin{equation}
 ds^2 >0 ; \qquad Timelike,
 \end{equation}
 \begin{equation}
 ds^2 =0 ; \qquad Null,
 \end{equation}
 and,
 \begin{equation}
 ds^2 <0 ; \qquad Spacelike,
 \end{equation}
Had we chosen a signature of $+2$ for the metric, the definition of
timelike metric would have been $ds^2 <0$ and so on. A timelike
worldline always (actually) remains so and no Lorentz  transformation can
change this intrinsic character [30].
 The same is true for other two kinds of
worldlines too.  This
notion is directly carried over to GTR because by POE, local laws of
physics are same as in STR. Further, as a matter of definition, both in
STR and GTR, the material particles follow (actually) timelike worldline,
photons and other massless particles follow null lines. Only fictitious {\em
tachyons} having imaginary mass parameter, follow spacelike line. Former two classes of worldlines are sometimes jointly
called ``non-spacelike'' worldlines  and, in general all causally connected
worldlines must be
nonspacelike [9, 30]. If a given worldline ever appears to change its intrinsic
character, it must be because,  either, the underlying
coordinate system is faulty i.e., it may harbor a singularity, or there may be
a {\em basic fault in the  formulation of the physical problem itself}.
 These definitions actually embody the
basic postulate of STR that ``nothing (associated with mass-energy) can
move faster than light'' and which gets embedded into GTR by virtue of POE.
\subsection{COMOVING FRAME}
By definition a comoving frame (COF) is one in which the test particle is
at rest.  This simple Euclidean or STR notion,  however, might get
slightly complicated in GTR, in the presence of  gravity.
Let us still start with the  same notion of a COF in which the
velocity of the fluid or particle is zero, $v_{com} \equiv 0$. The
components of the 4-velocity $u^i= dx^i/ds$, in the COF are:
\begin{equation}
u^\alpha (\rm COF) \equiv 0; \qquad u^0 (\rm COF) = (g_{00})^{-1/2}
; \qquad u_0 (\rm COF) = (g_{00})^{1/2}
\end{equation}
But unlike
in STR, the time measured by the same physical clock fixed in the COF, is
in general, not the proper time, i.e., in general $g_{00}^{cof} \neq 1$.
 The rate of ticking of
the same clock, measured locally, will change at various points along the
worldline. In otherwords, the COF is not, in general, a synchronized
coordinate system and the worldline in it is not a geodesic. But there is a
corresponding running frame in which, the time, recorded
locally, form a well ordered
causal sequence. This is called a Synchronized COF (SCOF) [30] in which
\begin{equation}
u^\alpha (SCOF) =0; \qquad
g_{00} (SCOF) \equiv 1; \qquad u^0 (\rm SCOF) = (g_{00})^{-1/2} =1
\end{equation}
Only, if the test particle is
under free fall, i.e., if the worldline is a geodesic, the COF time
becomes synchronized time [30].
Although, the value of $g_{00} \equiv 1$ in the SCOF, the value of the
various spatial components $g_{\alpha \beta} \neq -1$. If one constructs a
special COF in which the coordinate basis is orthonormal, it is called a
``proper frame''; most of thermodynamical quantities like pressure,
energy density, and temperature are measured in this frame.

On the other hand, the corresponding
 set of LIFs threading the COF worldline form a special
 synchronized coordinate system where
 \begin{equation}
 g_{00}^{LIF} =1; \qquad g_{\alpha \beta}^{LIF} = -1; \qquad g_{ik}^{LIF} =\eta_{ik}
 \end{equation}
which is called the Fermi Coordinate of the First kind

 However, obviously, the LIF is no COF   and
\begin{equation}
u^{\hat \alpha} =\gamma v^{\hat \alpha}; \qquad \gamma= \sqrt{1- v^2}; \qquad v^2
=v^{\hat \alpha} v_{\hat \alpha}
\end{equation}
where we use the notation of ``hat'' to use quantities measured by LIF of
the observer
and $v$ is the speed and
$\gamma$ is the Lorentz  Factor of the test particle measured in the LIF (now onwards, we take c=1).
We will elaborate on the nature of the vector $v^i$ later.
In GTR it is only this $\gamma$ which can be defined meaningfully. For
instance a $speed$ defined with respect to any distant observer has little
significance in a curved spacetime.
Thus
\begin{equation}
v\equiv {\hat v}; \qquad \gamma \equiv {\hat \gamma}
\end{equation}
If a true  COF is definable, the clock in it would be at a fixed spatial location,
and then we could write, for any range of $t$,:
\begin{equation}
ds= \sqrt{g_{00}^{com}} dt
\end{equation}
But, in general
\begin{equation}
ds= d\tau
\end{equation}
And since $d\tau$ must be positive, in a general comoving frame, we must have
\begin{equation}
g_{00}^{com} >0
\end{equation}
Had we taken a different signature of the underlying metric, this
condition, would have been $g_{00} <0$.
This shows that all general comoving frames must be characterized by a
definite sign (in this case +ve) of $g_{00}$. Here, by the term,
``general'' we imply that a comoving frame with $x^\alpha =constant$
exists, and which is always possible
when one is not presuming any specific
form of the metric coefficients and indeed measuring $t$ by a clock
rigidly fixed with the test particle.
This fact expresses the simple fact that, in
general, a COF is free of any
 coordinate singularity.
\subsection{General or Standard Frame}
In GTR, in principle, the choice of a coordinate system is arbitrary, and
one need not work only with comoving coordinates. We may call a general non-comoving coordinate system as the Dynamic
Standard Frame (DSF). In order to
maintain the general dynamic nature of the coordinates, it is
absolutely necessary to retain the general time dependent nature of the
corresponding metric coefficients $g_{ik}$. According to Landau and
Lifshitz (LL) [30] (pp. 234), the property of the COF, that $g_{00} >0$,
 should be characteristic
of any physically valid spacetime, and, if it is not so,  appropriate
coordinate transformation must be applied to restore this feature.
Suppose, the temporal coordinate in a DSF is $T$, then, without any loss
of generality, the proper time element following a worldline would be
\begin{equation}
d\tau = \sqrt{g_{TT}} dT
\end{equation}
In fact, we would see later that in order to describe the spherically symmetric
solutions, particularly, the solutions interior to the fluid, in principle
one can work with a ``standard'' coordinate system $R, T$ which is not comoving.
Although one usually works in the COF for the sake of mathematical
simplicity, in principle, one should be able to work with the standard
coordinate too provided one maintains the generality of the problem by
treating the metric coefficients to be unknown function of $R, T$ (and
does not put, beforehand, the simplified form of them valid in an empty
spacetime). We will realize at the end of this paper that most of the
so-called ``coordinate singularities'' actually result either because of
such tacit simplifications or because of the essential incorrect
formulation of the complicated physical problem in our eagerness to obtain
apparently ``approximate'' solutions and then to assert that the actual
complicated problem should ``approximately'' behave in the idealized manner.
\subsection{ Apparent Standard Frame}
It may appear that  the choice of a (correct) coordinate system is a trivial
matter because all one has to do is to assign a set of labels with an event.
In GTR it is not so because in the absence of an absolute (flat) spacetime
coordinates are meaningful only when they are organically associated with the
appropriate metric coefficients or the underlying spacetime geometry. In
GTR, the background geometry against which physical phenomenon are to be
analyzed itself is determined by the detailed physics. This does not,
however, mean, that we must have exact prior idea about the solution,
(i.e., the metric coefficients) before we proceed to solve it. On the other
hand, it only means that, we must not curb the generality of the problem
by presuming certain simplifications in the nature of the
metric coefficients. In principle, one should formulate the problem with
its full generality and then try to self consistently evaluate the metric
coefficients either analytically (which is possible rarely) or numerically.

But in some cases, we may be tempted, apparently justifiably, to presume a
specific form of the explicit metric coefficients and then proceed to
analyze the problem. On the basis of the presumed spacetime geometry
(metric coefficients), we set up a background grid of spacetime lines and
try to assign coordinates to the test particle; and let, in spherical symmetry,
these coordinates be $R, T$. Then we would be tempted to define a standard
frame by
collecting all $R=constant$ points. But since clearly, in such a case, we
have compromised with the generality of the problem we are not sure whether
$R, T$ really form a genuine dynamic standard coordinate. We shall dwell more on
this in Sec. 4.
\subsection{Proper Distance}
Following Landau \& Lifshitz (LL) [30],
a general metric may be rewritten in terms of its spatial part
$g_{\alpha \beta}$ as
\begin{equation}
ds^2=g_{\alpha \beta} dx^\alpha dx^\beta + 2g_{0\alpha} dx^0 dx^\alpha
+ g_{00} (dx^0)^2
\end{equation}
Then, it follows that,
the proper distance between two nearby events, as defined by LL [30]:
 \begin{equation}
 dl^2 = \gamma_{\alpha \beta} dx^\alpha dx^\beta
 \end{equation}
where
\begin{equation}
\gamma_{\alpha \beta} =\left(-g_{\alpha \beta} + {g_{0\alpha}
g_{0\beta}\over g_{00}}\right)
\end{equation}
For a static metric, the spacetime cross terms can be always made to
vanish by appropriate coordinate transformation, i.e., it is possible to have
$g_{0\alpha} =0$, and then, we will have
\begin{equation}
dl^2 =g_{\alpha \beta} dx^\alpha dx^\beta
\end{equation}
It may be reminded that finite proper distances can be meaningfully defined only if $g_{ik}$ are
independent of $x^0$ [30].
\subsection {Physical Speed v}
It is of utmost importance to be able to properly define the quantity,
$v$, which is the speed of the test particle or the fluid element measured
by LIF. By POE, locally, GTR reduces to STR (except for global
gravitational laws) and, in general, one can study physical problems even in the
presence of gravitation provided ordinary derivatives are replaced by
something called ``covariant derivatives''. The first requirement for this is to ensure that,
the frame one is working with indeed yields that the quantity $v \le 1$. If
not, one must choose another appropriate frame with $x^i \rightarrow
x'^i$ and then {\em reverify that} in the new coordinate
$v\rightarrow v'\le 1$. Given a certain coordinate system, the velocity
(vector) of the test particle with respect to LIF, also called {\em
physical velocity} is given by [30] :
\begin{equation}
v^\alpha = {dx^\alpha\over \sqrt{g_{00}} (dx^0 -g_\alpha dx^0)} ; \qquad g_{\alpha}= -{g_{0 \alpha}\over g_{00}}
\end{equation}
Under coordinate transformation, obviously, the components of $v^\alpha$
can change. The scalar $v^2$ formed from this vector is
\begin{equation}
v^2 =v^\alpha v_\alpha = {dl^2 \over g_{00} (dx^0 -g_\alpha dx^\alpha)^2}
\end{equation}
How can we verify that we have indeed worked out the correct expression
for $v^2$, a quantity, which is extremely important for the appreciation of this paper?
This is quite simple and following LL,  we just need to  rewrite the general
form of any metric in
 terms of the proper distance:
\begin{equation}
ds^2 = g_{00} (dx^0 -g_\alpha dx^\alpha)^2 - dl^2
\end{equation}
or,
\begin{equation}
ds^2 = g_{00} (dx^0 -g_\alpha dx^\alpha)^2 \left[ 1- {dl^2\over g_{00} (dx^0
-g_\alpha dx^\alpha)^2}\right]
\end{equation}
Now by noting that $g_{00} >0$ and the condition for a timelike
worldline is $ds^2 >0$, we can identify the last term
 within  the square bracket as $v^2$, in agreement with Eq. (3.22). Consequently,
 the local value of the Lorentz factor is $\gamma= \sqrt{1 -v^2}$. On the
other hand, the components of the
relevant 4-velocity
\begin{equation}
u^i = {dx^i\over ds} = {dx^i \over d\tau}
\end{equation}
has the components [30] :
\begin{equation}
u^{\alpha}= \gamma v^\alpha; \qquad u^{0} =\gamma ({g_{00}}^{-1/2} + g_\alpha v^\alpha)
\end{equation}
and for a static field, they are
\begin{equation}
u^{\hat \alpha}= \gamma v^{\hat \alpha}; \qquad u^{\hat 0} = \gamma
\end{equation}
For finding the components of the energy momentum tensor, it will be
useful later to remember that, in a general frame, we have
\begin{equation}
u^\alpha u_\alpha = \gamma^2 v^2= {v^2\over 1-v^2};\qquad u^0 u_0
=\gamma^2 = {1 \over 1- v^2}
\end{equation}
In a general frame,
for a static metric with $g_{0\alpha} =0$, and, we have the following simplifications
\begin{equation}
ds^2= g_{00} (dx^0)^2 -dl^2
\end{equation}
\begin{equation}
v= {dl\over \sqrt{g_{00}} dx^0}
\end{equation}
And, if there is a true COF or true DSF with $x^0$ as the appropriate time
following a worldline,  then,
 we can write
\begin{equation}
v= {dl\over \sqrt{g_{00}} dx^0} = {dl \over d\tau}
\end{equation}
It is extremely important to note that we have presented here the correct
expression for the {\bf 3-velocity or the physical velocity as measured by a
local observer} in a static field following Landau \& Lifshitz [30].
No other  standard textbook on GTR seem to contain this discussion,
and, many experts on GTR also seem to be confused about this important aspect.
One must note that, it is this $v$ defined by Landau\& Lifshitz which
{\bf appears in the Local Lorentz transformations, and one must have} $v <1$.
For the benefit of the readers we enclose few appendixes at the end of
this paper which contain the photocopy of the relevant portion from LL.
For the general static case, we also have
\begin{equation}
u^{\hat\alpha}= \gamma v^{\hat \alpha}; \qquad u^{\hat 0} =\gamma (g_{00})^{-1/2}
\end{equation}
a point, already mentioned before. And, in a static field, the conserved energy of the particle is
\begin{equation}
E_\infty= m g_{0i} u^i=m g_{00} u^0 = m\gamma \sqrt{g_{00}}
\end{equation}
where $E_\infty$ is the energy measured by the inertial observer,
$S_\infty$, situated at the spatial infinity in an asymptotically flat spacetime,
which is always possible for a static field.
The above relations are valid in arbitrary static field and even when the
particle is not in free fall.
\section {Spherically Symmetric Gravitational Field}
The most general form of a spherically symmetric metric, after appropriate
coordinate transformations, can be brought to a specific Gaussian form
[9, 30]:
\begin{equation}
ds^2 =  {\tilde A}^2(R,T) dT^2 - {\tilde B}^2(R,T) dR^2 -  R^2 (d\theta^2 +\sin^2 \theta d\phi^2)
\end{equation}
where ${\tilde A}$ and ${\tilde B}$  are to be determined self consistently for a given
problem. For the simplification of computation, it is customary to express:
\begin{equation}
{\tilde A}^2 = e^\nu; \qquad {\tilde B}^2 =e^\lambda
\end{equation}
Here $R$ is an appropriate radial marker (coordinate) and $T$ is the
coordinate time. A spherical symmetry precludes the possibility of the
existence of any rotation in the problem, and, hence, space-time
cross-terms should be absent in any physically meaningful choice of
coordinates. In other words, a spherically symmetric gravitational field is  a {\em
static} one and which is a subclass of a {\em stationary} field. A stationary but
non-static gravitational field is the one generated by a body with an
axial symmtry (it may be spinning about its symmetry axis). For the
spherical metric, if we consider
a $T = constant$ hypersurface and pick up a curve (circle) with $R=constant$ and
$\theta=\pi/2$, the value of the invariant line element would be
\begin{equation}
 ds = R d\phi
\end{equation}
The invariant circumference of the $T=constant$ circle would be $2 \pi
R$, and thus, we identify $R$ as the (invariant) circumference
variable. Thus for spatial measurements, we may construct a coordinate
system threaded by $R=constant$ circles and put clocks on this grid to
measure time $T(R)$. This coordinate system is, in general, dynamic
because the associated metric coefficients depend on $T$ and,
 in principle, can be used to
study any problem (including interior collapse) which (implicity) does not demand  violation of strict
spherical symmetry. Such a coordinate system is called the Standard
Schwarzschild Frame and, obviously, it is a dynamic frame.
\subsection{Comoving Schwarzschild Metric}
In many physical situations and, particularly,  for studying gravitational
collapse, it may be desirable to work in COF to explore the approach to
probable singular regions  and for the sake of simplicity. When we use a truly comoving coordinate, the
radial coordinate $r$ would remain fixed for a given fluid packet. The
most appropriate choice for $r$ would be to set it proportional to the
total (conserved) number of baryons $N(r)$ within a given radius. And,
then, by definition, the radial coordinate velocity must be zero. But, the
coordinate velocity measured by the ($R, T$) frame is non zero for a
dynamically moving fluid. And thus, in GTR, we must differentiate between
a comoving coordinate system $(r, t)$ and a background coordinate system
$(R, T)$. In Newtonian physics, even for studying comoving (Lagrangian)
description of the hydrodynamics, one uses the same  absolute background
spacetime even though one adds an appropriate convective term to the
respective temporal derivatives. But, in GTR, this is not possible. Thus,
for a comoving description of the physics, we need to modify the metric to
explicitly see the role of $r$ [9, 30]:
\begin{equation}
ds^2 = A(r,t)^2 dt^2 - B(r,t)^2 dr^2 - R(r,t)^2 (d\theta^2 +\sin^2 \theta d\phi^2)
\end{equation}
Again, it is customary to express the (new) metric coefficients in an
exponetial fashion with exponents $\nu_{co}$ and $\lambda_{co}$. To avoid
confusion with the Eq. (4.2), in this paper we shall avoid expressing the COF
metric coefficients in this exponential form.

A comoving frame, by definition, can be constructed in a region {\em filled
with mass-energy} and can   be naturally defined in the
 interior of any fluid.
Note that,  the very notion of a ``comoving frame''(COF)
 incorporates the fact that the worldlines of
material particles are timelike, which, as we have, emphasized before,
embodies the concept that there exists a $v <c$.
 If, on the other hand, somehow, it were possible to have
 (actual) $v>c$, the matter could not be brought to rest in any ``frame''
and the concept of a COF would break down.
 To appreciate this point, recall
that, for a bunch of photons, there can not be
any comoving frame. However, it does not pose any difficulty for photons, because, by
fiat, photons have the same speed in any frame, inertial or otherwise.
\subsection {Exterior Schwarzschild Metric (ESM)}
But, now, suppose we are going to describe a truly ``vacuum'' exterior
region around a spherical body solution, a spactime not containing a
single ``particle'' or photon.
The actual solution for the vacuum exterior region
 was found by Schwarzschild in 1916 :
\begin{equation}
ds^2 = g_{TT} dT^2 + g_{RR} dR^2 - R^2 (d\theta^2 +\sin^2 \theta d\phi^2)
\end{equation}
with
\begin{equation}
g_{TT} = \left(1 -{R_{gb}\over R}\right); \qquad g_{RR}= - \left(1-{R_{gb}
\over R}\right)^{-1}; \qquad R_{gb}= {2GM_b\over R c^2}
\end{equation}
Clearly, this metric is of a standard type described before.
and, obviously, the time parameter $T$ appearing here  is {\em not}
any comoving time measured by a clock attached to the test particle. On
the other hand, for this special standard frame, using Eq. (4.6), it is seen
that, $T$ is the time measured by a distant inertial observer $S_\infty$.

 Thus both $R$
and $T$ have some sort of absolute meaning in the External Schwarzschild
Metric (ESM). The inherent static nature of the resultant background
frame is evident from the fact that the metric coefficients $g_{RR}$ and
$g_{TT}$ neither depend on
 $T$ nor do they
 {\em involve} any $v$. Thus, formally, they are {\em like the interior metric
coefficients for a static relativistic star in hydrostatic equilibrium}.

In GTR, the existence
of a coordinate system  is inextricably linked with the (solution
of) the physical problem itself.
Any presence of any material
particle or photon in the external region would mean violation of the
condition behind a ``vacuum'' solution. Then, how, do we study the motion
of a test particle in this region. Of course, first we go to the ``test
particle'' assumption, i.e., we have a probe which has no associated mass energy.
But there is a much greater practical difficulty. Note that, for the
interior region, even if we choose to work with DSF, there is always a
background COF with coordinates $r, t$, where, $r$ can be taken as the
conserved number of baryons within $r=r\neq R$. But,
 in a region external to the fluid, there is no matter, and hence
there can not be any Schwarzschild comoving coordinate.  In this region, therefore, $R$
doubles up as the radial marker or the radial coordinate, $r= R$.

While we study
the motion of a ``test particle''  in the
exterior region, {\em we are actually studying
an idealized two-body problem}. In Newtonian gravity, a  problem
involving a massive spherical body $M$ and a tiny body $m$ gets reduced to
a two body problem because the exterior gravitational field of the massive body
can be represented by a point mass sitting at the center of the body. In GTR,
a similar thing happens (as long as we ignore the radiation emitted in
gravitational collapse) because of Birchoff's theorem. But the similarity
actually ends there. A Newtonian two body problem gets reduced
to a static problem because the total energy is conserved. But, in GTR,
this is not the case and if the speed of the test particle approaches $c$,
it can lose significant amount of ``energy'' by means of gravitational
radiation. Even if the total energy of the system comprising
 $M$ and $m$ is hardly changed, for the test particle, irrespective of the
smallness of its mass, emission of gravitational waves can severely affect
its dynamics and distort the spacetime structure in the vicinity of the particle.
 Essentially, a GTR two body problem is  not only a non-static
problem but it is a non-stationary problem too.
Therefore,  no static or even stationary metric can really handle it.
 However, for a
single test particle, assignment of a coordinate $r$ encompassing a fixed
number of baryons does not work, and it may appear that, the
$R=constant$ grid can really serve as the DSF
provided we allow $g_{RR}$ to be a function of $T$. But once we presume
the validity of the ESM, $g_{RR}$ ceases to be $T$ dependent.

We
may like to ignore the non-stationary nature of the problem in the
following way. Instead of a single test particle, we may consider a
spherically symmetric shell of
incoherent  test (dust) particles. In this way, we can restore
the spherical symmetry of the problem and pretend to suppress the emission
of gravitational waves. But then, again, for constructing a COF, we need to
differentiate between $r, t$ and $R, T$. Thus even when we ignore
the essential non-stationary nature of the problem, in the background,
there is a coordinate $r$ present in this case even though we are not able
to define it in a self-consistent manner.

Thus although $R$ can be used as a coordinate to analyze the problem the
 $R=constant$ lines  constitute only an Apparent Standard Frame (ASF).
We may think of taking ``snapshots'' of
the test particle at various epochs and pretend to construct a
corresponding ``comoving
frame'' by joining together these snapshots. This tantamounts to analyzing
the motion of the particle in the background of a lattice of {\em static}
$R=constant$ lines.
\section {Elements of Dynamics in Spherical Symmetry}
For any spherically symmetric gravitational field, the radial worldlines are
characterized by $d\theta =d\phi =0$. Also, it is a static field with
$g_{\alpha 0} =0$.
Further,
since actually, there always exists a real proper time and proper
distance, in a general fashion, we can define
\begin{equation}
v=  {dl\over d\tau}
\end{equation}
And, if we are working with a true COF, we will have $d\tau =A dt$ and
then we can write
\begin{equation}
v= {B(r,t) dr\over A(r,t) dt}
\end{equation}
For a true DSF, following the worldline, over the entire regime of
spacetime, we can extend the above definition to:
\begin{equation}
v= {\sqrt{-g_{RR}} dR\over \sqrt{g_{TT}} dT}
\end{equation}
Also we define a quantity,
$U$,
 the rate of change of the circumference radius
with the proper time following the radial worldline
(at a constant $r$). The specific physical significance of $U$ will be
discussed later.
\begin{equation}
U \equiv {dR \over d\tau}
\end{equation}
And, if we are working with a true COF
\begin{equation}
U \equiv {1\over A} {dR \over dt}= {{\dot R}\over A}
\end{equation}
In the true DSF, over the entire spacetime
\begin{equation}
U \equiv {1\over \sqrt{g_{TT}}} {dR \over dt}= {{\dot R}\over \sqrt{g_{TT}}}
\end{equation}
We further define another quantity, akin to curvature, which can be
most transparently defined  for a true COF and which expresses the
rate of change of the circumference radius with the proper distance:
\begin{equation}
\Gamma\equiv {1\over B} {dR\over dr} = { R'\over B} = {dR\over dl}
\end{equation}
Here an overdot denotes differentiation with the coordinate time and
a prime denotes differentiation with respect to the radial coordinate.
If we are working in a true DSF, where $g_{RR}$ is a function of $R, T$,
without any loss of generality,
the above definition degenerates into
\begin{equation}
\Gamma\equiv   {dR\over dl} = (-g_{RR})^{-1/2}
\end{equation}
This degeneracy of $\Gamma$ in the DSF, is allowed for any time dependent
problem, and also for a truly time independent problem, like the interior
mass-energy filled {\em hydrostatic} solutions of a star.

However, in an ASF, because of the inherent
contradiction,  $\Gamma_{ex} \neq (-g_{RR})^{-1/2}$  even though,
the general notion of $\Gamma$ actually exists.
For a truly COF or DSF, the three foregoing definitions can be combined to yield a very important relationship:
\begin{equation}
U\equiv \Gamma v
\end{equation}
which, because of its generality is valid in any region of the spacetime.
Now, we examine the validity of this equation  the external region.
\subsection{Exterior Schwarzschild Region}
Although, in these region it is not practically possible to work with a
true COF, yet, in the test particle assumption, and as long as we avoid
regions with coordinate singularity, without any loss of generality,
 we can define the physical speed of the particle in the $R, T$ coordinate:
\begin{equation}
v_{ex}= {\sqrt{-g_{RR}} dR\over \sqrt {g_{TT}} dt}
\end{equation}
Since, the definition of $U$ does not involve $r$, without any loss of
generality, we can write
\begin{equation}
U_{ex}= {1\over \sqrt{g_{TT}}} {dR\over dT}
\end{equation}
and although the definition of $\Gamma$ can not be properly translated to
the exterior region in the absence of our lack of knowledge of $r$ and $g_{rr}$,
we will see later, that, this concept nevertheless survives this
difficulty, and, there indeed exists a relationship
\begin{equation}
U_{ex}= \Gamma_{ex} v_{ex}
\end{equation}
\section {Schwarzschild Singularity}
The notion of BHs  in GTR can be traced to Schwarzschild's discovery of
the spherically symmetric external solution in 1916 [28], described by Eq.
(4.6). On the other hand, it was a source of relief to Einstein [1] to
find that bodies in hydrostatic equilibrium can not be squeezed inside $R
<R_g$. But dynamically collapsing bodies should reach $R=R_g$ boundary,
and, once it happens, it would require infinite amount of reverse
acceleration to hold the particle fixed at $R=R_g$ and therefore, there
can not be any hydrostatic equilibrium at this stage. So the body must
continue to collapse till it is ``crushed'' to infinite density at the
central singularity $R=0$. Very crudely, this is the essential idea behind
the assertion that existence of BHs (or other singularities) is an
inevitable prediction of GTR. In other words, the existence of BHs or
other singularities are considered a great triumph for GTR even though
Einstein  was uncomfortable to accept the existence of probable
singularities in his theory. The aim of this paper is to critically
examine these issues by bringing out some fine points which have always
been overlooked. We start with the dynamics of a freely falling test
particle in the exterior SM.  Since we are interested only in radial
motions, we may, simplify the external SM metric as
\begin{equation}
ds^2= g_{TT} dT^2 + g_{RR} dR^2
\end{equation}
For a free particle, the Lagrangian is given by [9, 47]:
\begin{equation}
L= {1\over2} p^i p_i = - {m^2\over 2}
\end{equation}
The components of the $p^i$ in the ($R, T$) frame are:
\begin{equation}
p^T = {dT\over d\tau}; \qquad p_T = g_{TT} p^T= \left( 1-
{2GM\over R}\right) {dT\over d\tau}
\end{equation}
and
\begin{equation}
p^R=  {dR\over d\tau}; \qquad p_R = g_{RR} p^R= \left(1-
{2GM\over R}\right)^{-1} {dR\over d\tau}
\end{equation}
It is important to note here that since  $T$ is not a comoving time,
and neither is $R$ the true comoving coordinate, $p^T$ and $p^R$ {\em are not
the components of $p^i$ measured in  LIF}, and, on the other hand,
they may be related to quantities measured in $S_\infty$ because $T$ is
the proper time measured in this frame. This point will become  clear soon.
 We have already discussed that,  the actual
radial component of $p^i$ measured in LIF is
\begin{equation}
p^{\hat R} = \gamma v^{\hat R} =\gamma v
\end{equation}
In terms of $p^T$ and $p^R$, the Lagrangian looks like
\begin{equation}
2L= - m^2 \left( 1-
{2GM\over R}\right) \left({d T \over d\tau}\right)^2+ m^2 \left(1-
{2GM\over R}\right)^{-1} \left({d R\over d\tau}\right)^2 = -m^2
\end{equation}
With respect to the LIF, where, $p^{\hat i} =p_{\hat i}$, the same Lagrangian is
\begin{equation}
2L = -(p^{\hat T})^2 + (p^{\hat R})^2 = -m^2
\end{equation}
In order that Eqs. (6.6) and (6.7) are satisfied for arbitrary $R, T$, we must have
\begin{equation}
p^{\hat T} = m (1- 2G M/R)^{1/2} {dT \over d\tau} =m (1- 2G M/R)^{1/2} p^T
\end{equation}
\begin{equation}
 p^{\hat R} = m
(1-2GM/R)^{-1/2} {dR\over d\tau}= m (1- 2G M/R)^{-1/2} p^R
\end{equation}
Since $T$ is a cyclic coordinate, the corresponding Euler -Lagrange
equation yields
\begin{equation}
p_T=  m \left( 1-
{2GM\over R}\right) {d T\over d\tau} = m \left( 1-
{2GM\over R}\right) p^T= {\rm constant} =E_\infty
\end{equation}
where, we can identify $E_\infty$ as the energy measured by the inertial observer, $S_\infty$, sitting
at $R=\infty$, in an asymptotically flat spacetime.
 But, by Eq. (6.8), the energy measured in the LIF is
\begin{equation}
{\hat E} = p^{\hat T} = m (1- 2G M/R)^{1/2} p^T
\end{equation}
By combining the two foregoing Eqs., we have
\begin{equation}
E_{local}= {\hat E} = p^{\hat T} = m (1- 2G M/R)^{1/2} E_\infty
\end{equation}
Using the foregoing Eq. in Eq. (6.8), we obtain an
important relation:
\begin{equation}
{\tilde E}^2 = 1 + \left({dR \over d\tau}\right)^2 - {2G M\over R}
\end{equation}
where
\begin{equation}
{\tilde E} = {E_\infty \over m}
\end{equation}
is the conserved energy per unit rest mass as measured by $S_\infty$.
By recalling that
\begin{equation}
p^R= m {dR\over d\tau} \equiv m U_{ex}
\end{equation}
 we reexpress Eq.(6.13) as
\begin{equation}
{\tilde E}^2 = 1 + U_{ex}^2 - {2G M\over R}
\end{equation}
The radial velocity of the particle  measured in LIF is
\begin{equation}
v=v^{\hat R} ={p^{\hat R} \over p^{\hat T}} = {p^R \over E_\infty} =
{U_{ex}\over {\tilde E}}
\end{equation}
Or, we have
\begin{equation}
U_{ex} = {\tilde E} v_{ex}
\end{equation}
Comparing this foregoing Eq.  with Eq. (5.12),
we confirm that, in the ESM,
a $\Gamma_{ex}$
exists and is given by
\begin{equation}
\Gamma_{ex} = {\tilde E} ={E_\infty \over m}; \qquad \Gamma_{ex} \neq (-g_{RR})^{-1/2}
\end{equation}
In terms of this identification of $\Gamma_{ex}$, we may rewrite Eq. (6.16) as
\begin{equation}
{\Gamma_{ex}}^2 = 1 + U_{ex}^2 - {2G M\over R}
\end{equation}
If the particle is released at  $R=R_\infty$, we have
\begin{equation}
{\tilde E}^2 = \Gamma_{ex}^2= 1- {2G M\over R_\infty}
\end{equation}
Physically, this definition demands that $\Gamma_{ex}$ is always finite and
\begin{equation}
\Gamma_{ex} \le 1
\end{equation}
One may, however, imagine that if the particle was injected into the
gravitational field with a $v \neq 0$ at $R= \infty$, it would be possible
to have a value of $\Gamma_{ex} > 1$. Since, at $R=\infty$, there is no
gravitational field, the existence of such an initial condition would
imply the existence of external fields (like electromagnetic field). And
this is not allowed by the present problem. However, in STR, where there is
no gravity, the motions are necessarily because of  mechanical or other
forces, and it is possible to have an initial velocity gteater than zero
at infinity.
Now, by using Eqs. (6.17), (6.20) and (6.21), it follows that,
\begin{equation}
U_{ex}^2 = 2GM \left( {1\over R} -{1\over R_\infty}\right)
\end{equation}
and
\begin{equation}
v_{ex}^2 = {2GM \left( {1\over R} -{1\over R_\infty}\right) \over 1- {2GM\over R_\infty}}
\end{equation}
For $R_\infty = \infty$, we obtain,
\begin{equation}
{\tilde E} = \Gamma_{ex}= 1
\end{equation}
and
the physical velocity assumes the familiar
Newtonian form:
\begin{equation}
v_{ex}= \left({2GM\over R}\right)^{1/2} = \left({ R_g\over R}\right)^{1/2} c
\end{equation}
Therefore, (for a finite value of $R_g$) at $R=R_g$, we find, $v_{ex} =c$
and this (apparent) fact  is, somewhat
confusingly, used to call the surface with $R=R_g$ as the ``null
surface''. It is confusing because, as we will see below that it is
believed that $v_{ex}$ behaves anomalously at $R=R_g$ because of
coordinate singularity. And, if the coordinate singularity is removed by
choosing an appropriate new coordinate system, the (correct) worldline
would really be timelike everywhere. Further, the value of $v_{ex} >c $ for $R <R_g$ and
thus this region appears to be ``spacelike''. We repeat that such
 nomenclatures are unwarranted and
confusing because, in the new coordinate system, all (actual) worldlines
are expected to be timelike.

The (proper) radial acceleration acting on the particle and as is measured in LIF can be found to be
\begin{equation}
f=f^{\hat r} = {GM \over R^2 (1- R_g/R)^{1/2}}
\end{equation}
The fact that, $f \rightarrow \infty$ as $R \rightarrow R_g$ expresses
the fact that no amount of Lorentz boost can hold a particle fixed in a
COF characterized by $R= constant$. The singularity in the value of $R= R_g$
(for a finite $R_g$) clearly points out that the applicability of the $(R, T)$ coordinate
system {\em must be restricted} to $R > R_g$. On the other hand,
 the tidal acceleration measured in
the LIF $\sim GM/R^3$, however, remains finite even for $R \le R_g$ (if $R_g$ is
finite). The various components of the Rimmenian curvature tensor
$R_{ikml}$ are
directly related to this tidal acceleration, and the fact that they remain
finite, led to the idea that the Schwarzschild Singularity
is a mere coordinate singularity. Further, one may construct several
scalars by contracting various components of the curvature tensor, the
simplest of which is the ``scalar curvature'' of spacetime
\begin{equation}
C= R^{ikml} R_{ikml}= {12 R_g\over R^6}
\end{equation}
All such scalars too remain finite at $R=R_g$ (if $R_g$ is finite).
It is believed that this singularity
 must vanish (even when we ignore the emission of gravitational
radiation), in an appropriate
new coordinate  valid either for the entire spacetime (Kruskal coordinate)
or atleast for $R\le R_g$ (Lemaitre Coordinate). There is a subtle
inconsistency in this theoretical stand; because once we reject the ESM for the
$R \le R_g$ region and even when some authors conclude that, in the
interior region (called $T$-region), $R$ becomes (actually) spacelike and
$T$ becomes (actually) timelike, the components of the curvature tensor,
expressed in the same ($R, T$) coordinate would have little physical
meaning although, the scalars are meant to be unchanged. For a true self consistency, one ought to recalculate the
curvature components in whatever new (correct) coordinates one adopts.
Further, one should also try to find $v$ {\em independently} in the new
coordinate to ensure that it is indeed $< 1$. When, we say, ``independently'',
we mean that, it should evaluated for a material particle. If we use a
null worldline condition $ds^2=0$ for such a derivation, we would be
presetting $v=1$ in Eq. (3.24), and then, as a tautology, any coordinate system,
physically valid or invalid, will always reproduce $v=1$.
\subsection{Lemaitre Coordinate}
One of the basic difficulties with the external SM was that it simply does
not admit of a seperate radial coordinate $r$ and the circumference
coordinate $R$ had to be used for marking radial seperation as well.
Whereas this behavior is perfectly alright for a fluid in hydrostatic
equilibrium, it prohibited the use of a truly dynamic COF. This
undesirable situation may be remedied by explicitly introducing a radial
coordinate $r$ rigidly attached to the moving particle, and, the price one
pays for it is that the simple external SM ceases to be valid; the new
metric coefficients would explicitly involve the time $t$ recorded in the
true COF. However, for a freely falling particle, $g_{00}^{ff}
\equiv 1$ and thus $t$ becomes synonymous with the true (synchronous)
proper time $\tau$. It was Lemaitre [18], who, suggested a COF for the region
$R \le R_g$ [30] :
\begin{equation}
ds^2 = d\tau^2 + g_{rr} dr^2 -R(r,\tau)^2 (d\theta^2 +\sin^2 \theta d\phi^2)
\end{equation}
where
\begin{equation}
- g_{rr} = \left[{3\over 2 R_g} (r - \tau)\right]^{-2/3}
\end{equation}
and
\begin{equation}
R = \left[{3\over 2} (r -\tau)\right]^{2/3} R_g^{1/3}
\end{equation}
Here, $R=R_g$ corresponds to $r-\tau =(2/3) R_g$ and the central singularity $R=0$
corresponds to $r-\tau =0$. Evidently, the new metric is regular
everywhere except at $R=0$ where the physical singularity is present. The
total spacetime is now analyzed by a two piece  coordinate system, (1)
External SM for $R > R_g$ and (2) Lemaitre Coordinate for $ R\le R_g$.
And, in this hybrid coordinate, the entire spacetime  (except $R=0$
point) is believed to be well behaved with {\em all geodesics as
timelike}, as they must be by  the postulations of GTR. Here it is
worthwhile to point out that the Lemaitre solution, like the external SM
solution is a {\em vacuum} solution and describes the vacuum region inside
the event horizon. Thus, {\em it does not have a direct applicability even for the
collapse of a ``dust'' because all collapse solutions necessarily involve
presence of mass-energy}. Further, the strict dynamic  evolution of
matter possessing pressure can not be described by a free fall.
\subsection{ Kruskal Coordinate}
It is desirable to describe the entire spacetime corresponding to a {\em
vacuum} Schwarzschild solution describing free fall of a test particle by
means of a single regular coordinate system. It was achieved by Kruskal [31]
(1960) in terms of a coordinate system, $(r_*, t_*)$, possessing rather unusual properties:
\begin{equation}
r_*^2 - t_*^2 = K^2 \left( {R\over R_g} -1\right) exp \left({R\over R_g}\right)
\end{equation}
where $K^2$ is an arbitrary constant, and is not to be confused with the
$K$ introduced in Sec. 2. These coordinates also satisfy the condition
\begin{equation}
{2 r_* t_* \over r_*^2 + t_*^2} \equiv \tanh \left ({t\over R_g}\right)
\end{equation}
The metric now looks like
\begin{equation}
ds^2={4 R_g \over R K^2} exp \left(- {R\over R_g}\right) (dt_*^2 - dr_*^2) -
R(r_*, t_*)^2 (d\theta^2 + d\phi^2 \sin^2 \theta)
\end{equation}
So, here, we have,
\begin{equation}
- g_{r_* r_*} =g_{t_* t_*} = {4 R_g \over R K^2} exp \left(- {R\over R_g}\right)
\end{equation}
Here, the central singularity corresponds to
\begin{equation}
r_*^2 -t_*^2 \rightarrow - K^2
\end{equation}
and the event horizon corresponds to $r_* \rightarrow t_*$.
Although, this metric appears to be perfectly regular except at $R=0$, it
is important to remember that $r_*$ and $t_*$ are
 {\em not comoving coordinates},
as is evident from the fact that $g_{t_* t_*} \neq 1$ even though the
particle is under free fall.
Being a one-piece coordinate system, it could not have been comoving, because, while it is possible to describe the interior region
$R \le R_g$ by a truly comoving coordinate $r$,  it
is not possible to do so in the exterior region.
And, needless to say that {\em it also describes a
vacuum spacetime and has therefore no direct relevance for the
collapse problem}.

\section{Formulation of the Collapse Problem}
Although, our central result would not depend on the details of the numerous
 equations involved in the GTR collapse problem, yet,
 for the sake of better appreciation by the reader, we shall outline the
general formulation
of the GTR spherical collapse problem, and refer, the reader to the respective original papers for
greater detail. The general formulation for the GTR collapse of a perfect
fluid, by ignoring any emission of radiation, i.e, for adiabatic collapse,
 was given by Misner and
Sharp [10] and May and White [32]. We may start with the Einstein
equation itself:
\begin{equation}
R_{ik} = 8\pi G \left( T_{ik} - {1\over 2} T\right)
\end{equation}
where the energy momentum stress tensor for a perfect fluid is
\begin{equation}
T_{ik} = (\rho +p) u_i u_k - p g_{ik}; \qquad T^i_k =(\rho +p)u^i u_k - p \delta^i_k;
 \qquad T=T^i_i = 3p- \rho
\end{equation}
Here $p$ is the isotropic pressure (in the proper frame)
and the total energy density of the fluid
in the same frame (excluding any contribution from global
self-gravitational energy) is
\begin{equation}
\rho =\rho_0 + e
\end{equation}
where $\rho_0 = m n$ is the proper density of the rest mass, $n$ is the
number density of the baryons in the same frame (one can add leptons too),
 and $e$ is the proper
internal energy density.
Here $R_{ik}$ is the contracted (fourth rank) Rimmenian curvature tensor
$R^i_{jkl}$, and, is called the Ricci tensor. In terms of the Christoffel
symbols
\begin{equation}
\Gamma^i_{kl} = {1\over 2} g^{im} \left({\partial g_{mk}\over \partial
x^l} + {\partial g_{ml}\over \partial x^k} - {\partial g_{kl}\over
\partial x^m}\right)
\end{equation}
the components of the Ricci tensor are
\begin{equation}
R_{ik} = {\partial \Gamma^l_{ik}\over \partial x^l} - {\partial
\Gamma^l_{il}\over \partial x^k} + \Gamma^l_{ik} \Gamma^m_{lm} -
\Gamma^m_{il} \Gamma^l_{km}
\end{equation}
By using Eq. (3.28), it is illustrative here to see that the components of $T^i_k$ in
any spherically symmertic coordinate system where $i =0, 1,2,3$:
\begin{equation}
T^1_1= (\rho +p) \gamma^2 v^2 -p; \qquad T^2_2=T^3_3=-p; \qquad
T^0_0=\gamma^2 (\rho+p) -p; \qquad T^1_0=T^0_1 ={\gamma^2 v^1\over \sqrt{g_{00}}}
\end{equation}
In the COF, $v^\alpha =0$, and the algebra is greatly simplified
because $T^i_k$ is diagonal too:
\begin{equation}
T^r_r =T^\theta_\theta=T^\phi_\phi =- p; \qquad T^0_0 =\rho
\end{equation}
It may be noted here that, in general, the components of $T^{ik}$ explicitly
involve $r, \theta, \phi$ in the COF, for instance $T^{rr} = p r^{-2}$,
and it is only the components of the mixed tensor $T^i_k$ which assume
very simple forms.
 One also requires to use the local energy momentum
conservation law:
\begin{equation}
T^i_k; k =0
\end{equation}
where a semicolon, ``;'', denotes covariant differentiation:
\begin{equation}
T^i_ k; l = {\partial T^i_k \over \partial x^l} -\Gamma^m_{kl} T^i_m +
\Gamma^i_{ml} T^m_k = 0
\end{equation}
 In the LIF, this law gets
simplified to
\begin{equation}
{\partial T^{{\hat i} {\hat k}}\over \partial x^{\hat k}} =0
\end{equation}
i.e., the law reduces to its STR form by virtue of POE. However, for our
purpose, we would not really require to invoke any explicit energy
momentum conservation (local) law. One has to supplement these equations
with the equation for continuity of baryon number :
\begin{equation}
(nu^i); i =0
\end{equation}
Now, after considerable algebra, in the COF,
the Einstein equations become [32] :
\begin{equation}
(R^0_0):~~~~~~ 4\pi G \rho R^2 R' = {1\over 2} \left( R+ {R {\dot R}^2\over
A^2} -{R R'^2\over B^2}\right)'
\end{equation}
\begin{equation}
(R^r_r): ~~~~~ 4\pi G p R^2 {\dot R} = -{1\over 2} \left( R +{R {\dot R}^2\over
A^2} -{R R'^2\over B^2}\right)^.
\end{equation}
\begin{equation}
(R^{\theta}_{\theta}, R^{\phi}_{\phi})~~~~: 4\pi G (\rho+p)R^3=\left(R+ {R{\dot R}^2\over
A^2}-{RR'^2\over B^2}\right)+{R^3\over AB} \left[\left({A'\over
B}\right)'-\left({{\dot B}^.\over A}\right)\right]
\end{equation}
\begin{equation}
(R^r_0, R^0_r): ~~~~~~ 0 = {A'{\dot R}\over A} + {{\dot B} R'\over B} -
{\dot R}'
\end{equation}
Further, if we define a new function
\begin{equation}
M(r,t) = 4\pi \int^r_0 \rho R^2 dR = 4\pi \int^r_0 \rho R^2 R' dr
\end{equation}
the $R^0_0$- field Eq. (7.12) can be readily integrated to
\begin{equation}
R + {R {\dot R}^2 \over A^2} - {R R'\over B^2} = 2 GM
\end{equation}
where the constant of integration has been set to zero because of the
standard central boundary condition
$M(0,t) =0$.
Here the coordinate volume element of the fluid is $dV =4 \pi R^2 dR$.
If we move to the outermost boundary of the fluid situated at a fixed
$r=r_b$, and demand that the resultant solutions match with the
exterior Schwarzschild solution, then we would be able to identify
$M(r_b)$ as the function describing the total (gravitational) mass  of the
fluid {\em as measured by the distant inertial observer} $S_\infty$.  For
an interpretation of $M(r, t)$ for the interior regions, we first, recall
that the element of proper volume is
\begin{equation}
d {\cal V} = 4\pi R^2 dl= 4 \pi R^2 \sqrt {-g_{rr}} dr = {4\pi R^2 dR A\over
R'}  = {dV \over \Gamma}
\end{equation}
so that
\begin{equation}
 M(r,t) =  \int^r_0 \rho dV  = \int^r_0 (\Gamma \rho) d {\cal V}
\end{equation}
For a free particle in the exterior Schwarzschild Metric, we found that
$\Gamma_{ex}={\tilde E}$ is the conserved energy per unit proper mass. And
this suggests that $\Gamma \rho$ is the energy density measured by $S_\infty$.
We have already inferred that $M_b$ is the total mass energy of the fluid
as seen by $S_\infty$. Then for a self-consistent overall description,
 we can interpret that, in general, $M(r,t)$ is the mass-energy within
$r=r$ and as sensed by $S_\infty$.

Now using our earlier notation, $U= {\dot R\over A}$ and $\Gamma={R'\over B}$,
we reframe Eq. (7.17) as
\begin{equation}
\Gamma^2 = 1 + U^2 - {2GM\over R}
\end{equation}
Thus, even for the internal solutions, we recover a relation exactly
similar to what was obtained for a test particle in an exterior region
(Eq. 6.20). Although, even at this stage we could have presented our simple exact
central result, for a deeper insight into this problem, we will defer our
simple derivation. Further using a compact notation
\begin{equation}
D_{t}= {1\over A} {d\over dt}\mid_{r=r}; \qquad D_{\rm r}= {1\over
B}{d\over dr}\mid_{t=t}
\end{equation}
the major adiabatic collapse equations turn out to be
[10, 32]
\begin{equation}
D_{\rm r}M=4\pi R^2 \rho \Gamma
\end{equation}
\begin{equation}
D_{\rm t}M= -4\pi R^2 p U
\end{equation}
\begin{equation}
D_{\rm t} U=-{\Gamma\over \rho+p} \left({\partial p\over \partial R
}\right)_{\rm t} -{M+4 \pi R^3 p \over R^2}
\end{equation}
\begin{equation}
D_{\rm t} \Gamma=-{U \Gamma \over \rho +p}\left({\partial p \over \partial
R}\right)_{\rm t}
\end{equation}
An immediate consequence of the last equation is that, if we assume a
$p=0$ EOS, $\Gamma$ will be time independent
\begin{equation}
\Gamma (r, t)= \Gamma (r)
\end{equation}
and for a fixed comoving coordinate $r$, $\Gamma$ would be a constant.
Further, the Eq. (7.23) shows that for $p=0$, we also have
\begin{equation}
M (r, t)= M (r) =constant
\end{equation}
After the formalism for GTR collapse equations were developed in the
sixties, the new arguments favouring BH formation go like this [22]:

For collapse, $U<0$, and from Eq. (7.23) it follows that $M(r,t)$ increases
monotonically within the fluid; of course, it remains constant, for
adiabatic collapse, for $r\ge r_b$, the outer boundary. Because of this,
the $M/R^2$ term in Eq. (7.24) increases more rapidly than it would in
the Newtonian theory where it remains constant. Then Eq.(7.25) would
suggest that $\Gamma$ {\em decreases monotonically} instead of remaining
constant in a Newtonian case (note that for a dust too, $\Gamma$ remains
constant in a Newtonian fashion). Both of these relativistic effects are
supposed to cause the collapse process monotonic and accelerate faster
than the Newtonian case [22].  Here note that unless we presume that
$\partial p/\partial R$ can be positive,  a negative value of $\Gamma$ in
Eq.(7.25) would demand that $\Gamma$ increases with time implying $\Gamma$
{\em would tend to be back to zero}.

Having laid this basic formalism for adiabatic collapse, we would move to
the simplified case of  ``dust collapse'' which happens to be a special case
of adiabatic collapse.

\section { Dust Collapse}
There is no way we can ever think of exactly solving the adiabatic
collpase equations for a real fluid, i.e., one having pressure. Further,
this idea of adiabaticity would break down as soon as the fluid starts to
contract because of the Kelvin-Helhmoltz energy liberation. Even if we
consider the fluid to be degenerate and at $T=0$ to start with,
gravitational contraction would keep on heating it up unless it acquires
an effective adiabatic index $\gamma =4/3$. On the other hand, we may
feign to ignore the role of any temperature in the fluid by artificially
assuming a polytropic EOS, $p\propto \rho^\gamma_t$ even when the gas is non-
degenerate. But the value of $\gamma_t$ will keep on evolving and it is not
possible to find any unique solution for the entire range of $p$ and
$\rho$ even by any numerical means. Depending on the inevitable
hidden assumptions
made, it may be possible to obtain any number of of solutions (by any
number of authors) and none of which
may have to do much with the actual complicated physics of atomic and
nuclear matter at arbitrary high density and pressure. And, the only way,
one may hope to obtain an exact or near exact solution, at the cost of the
actual thermodynamics, is to do away with the EOS, i,e., to set $p\equiv
0$ even when $\rho \rightarrow \infty$. Since the early days of GTR, dust solutions
were used in cosmology on the plea that in the present epoch, for the
cosmic matter $p \ll \rho$. This seems to be reasonably justified in the
cosmological context and, in any case, the assumption of perfect isotropy
and homogeneity of cosmic mass-energy distribution yields a metric
(Robertson Walker metric) which is the same as the one due to a
homogeneous dust [9].

 To justify the
``dust'' assumption, traditionally, the argument goes like the following:
during dynamic collapse, the star would any way collapse within its event
horizon, tacitly assuming that, $Q \ll M_i c^2$, so that the value of
$R_{gb}$ remains fixed by the initial conditions. Then,
 once the fluid has collapse within the event horizon, and if one
(incorrectly, for a finite $R_g$)  uses the
idea that at the Schwarzschild Singularity,  the gravitational
attraction would be infinitely strong, a dust solution may be justified
and give qualitatively correct result. Nevertheless, we recall that, in the context of a
finite $R_g$, all that the Schwarzschild Singularity meant that, there can
not be any hydrostatic equilibrium and a dynamical collapse would be
inevitable (not that the actual tidal acceleration would be infinite, if $R_g$
is really finite).
Thus we find that, the dust solution may not represnt the qualitative behaviour
associated with the collapse of a real fluid, if $R_g$ is really finite.

The GTR collapse problem for the inhomogeneous dust was first studied by
Tolman [39] and then by Datt [5], Bondi [20] and LL [30]. As mentioned before, for a dust,
the COF  itself is the synchronous frame so that COF $t\equiv \tau$, the
proper time. By setting $A= 1$,  the COF metric for a dust is
\begin{equation}
ds^2 = d\tau^2 - B^2(r, \tau) dr^2 - R^2 (r, \tau) (d\theta^2 +
\sin^2 \theta d\phi^2)
\end{equation}
By using this COF coordinate Tolman [39] obtained the following general
 equations
which are {\em valid for the entire spacetime in the interior of the
fluid} and
which is a necessary property of all COF coordinates or appropriate DSF coordinates:
\begin{equation}
B^2 =- g_{rr} = {R'^2 \over 1+f(r)}
\end{equation}
And this equation holds only if
\begin{equation}
1 +f >0
\end{equation}
 One also obtains
\begin{equation}
{\dot R}^2 = f(r) + {F(r) \over R}
\end{equation}
\begin{equation}
8 \pi G \rho = {F'(r) \over R' R^2}
\end{equation}
However if we really want to see any explicit behaviour of the metric
coefficients as a function of time, one has to further simplify the
problem by assuming the dust to be homogeneous:
\begin{equation}
\rho = {\rm constant}; \qquad r<r_b; \qquad \rho =0; \qquad r> r_b
\end{equation}
We study below the resultant solutions following the pioneering work of
Oppenheimer and Snyder [26].
\subsection {Oppenheimer -Snyder Solution}
Since the OS solutions are the only (asymptotic and near exact) solutions for
the GTR collapse, and are believed to explicitly show the formation of
``trapped surfaces''  and ``event horizon'', it is
extremely important to critically reexamine them.
Like Tolman [39], OS
initially worked in the COF, but, then, to match the internal solutions
with the external ones, eventually shifted to the  (non comoving) DSF
 involving $R, T$. The important point to remember here that OS used a real Dynamical
Standard Frame and not the Apparent SF involving only the External SM (ESM).
Thus, their approach was perfectly justified from this view point, and the
metric coefficients listed below, in principle, represent the state of the
fluid at an arbitrary stage of collapse including $R \rightarrow 0$.
Without giving the details of the actual mathematical manipulations, we
shall simply present their key equations. By matching the internal
solutions with the exterior ones they obtained a general form of the
metric coefficients and also a relation between $T$ and $R$ which
is valid for the entire range of $ \infty > R >0$ :
\begin{equation}
g_{TT} =e^\nu =  \left[ (dT/d\tau)^2 (1-U^2)\right]^{-1}
\end{equation}
\begin{equation}
-g_{RR} = e^\lambda =(1-U^2)^{-1}
\end{equation}
and,
\begin{equation}
T = {2\over3} R_{gb}^{-1/2} (r_b^{3/2} -R_{gb}^{3/2} y^{3/2}) -2 R_{gb} y^{1/2} +
R_{gb} \ln {y^{1/2} +1 \over y^{1/2} -1}
\end{equation}
where
\begin{equation}
y \equiv{1\over 2} \left[ (r/r_b)^2 -1\right] + {r_b R\over R_{gb}  r}
\end{equation}
It is the above Eq. (8.9) which corresponds to Eq. (36) in the OS paper.

It is important to note that for the outermost surface $r=r_b$, we have
\begin{equation}
y=y_b = {R_b\over R_{gb}}
\end{equation}
OS also showed that the relation between $T$ and $\tau$ is determined by
\begin{equation}
F \tau + r^{3/2} = R^{3/2}
\end{equation}
where,
\begin{equation}
F = - (3/2) R_{gb}^{1/2} (r/r_b)^2 ; \qquad r\le r_b
\end{equation}
So, for the outer boundary, we have
\begin{equation}
\tau = {2\over 3} { r^{3/2} - R^{3/2} \over (r/r_b)^2 R_{gb}^{1/2}}
\end{equation}
According to OS, in the limit of large T, one can write
\begin{equation}
T \sim -R_{gb} ~\ln \left\{ {1\over 2}\left[ \left({r \over r_b}\right)^2 -3\right] + {r_b
\over R_{gb}} \left( 1- {3 R_{gb}^{1/2} \tau \over 2 r_b^2}\right)\right\}
\end{equation}
The last term of the above equation contains a typographical
error and the corrected form should be
\begin{equation}
T \sim -R_{gb}~ \ln \left\{ {1\over 2}\left[ \left({r \over r_b}\right)^2 -3\right] + {r_b
\over R_{gb}} \left( 1- {3 R_{gb}^{1/2} \tau \over 2 r_b^{3/2}}\right)\right\}
\end{equation}
From the foregoing equation, they concluded that, ``for a fixed value of $r$
as $T$ tends toward infinity, $\tau$ tends to a finite limit, which
increases with $r$'' . They did not specify any form of this ``finite
limit'' for $\tau$. Actually by using the Eq. (8.14) , we can see more
transparently that the finite limit is $\propto R_{gb}^{-1/2}$.
Also, it follows that
\begin{equation}
e^{-\lambda} = 1 -(r/r_b)^2 \left\{e^{-T/R_{gb}} +{1\over 2} \left[3 - (r/r_b)^2\right]\right\}^{-1}
\end{equation}
and
\begin{equation}
e^\nu = e^{\lambda- 2 T/R_{gb}}
\left\{ e^{- T/R_{gb}} + {1\over 2} \left[3-(r/r_b)^2 \right]\right\}
\end{equation}

 Further, they
pointed out that for very large $T$, the metric coefficients behave in the
following way:
\begin{equation}
e^{\lambda} \rightarrow \infty; ~r<r_b; \qquad e^\lambda
={\rm finite}; ~for~ r=r_b
\end{equation}
and
\begin{equation}
e^\nu \rightarrow e^{- 2 T/R_{gb}}; \qquad r< r_b; \qquad e^\nu \rightarrow
e^{-T/R_{gb}}; \qquad r=r_b
\end{equation}
 Note that the Dynamic Schwarzschild frame
used by OS is a perfect legitimate frame for studying the internal
collapse, and this dynamic nature is reflected in the explicit temporal
dependence of the metric coefficients. And therefore one should not face
any difficulty in exploring the $R=0$ limit and OS indeed
 tried to probe the $R \rightarrow 0$ limit
 ``For $\lambda$ tends to a finite limit for $R \le R_{gb}$ as $T$
approaches infinity, and for $R_b=R_{gb}$ tends to infinity. Also for $R \le
R_{gb}$, $\nu$ tends to minus infinity.'' Actually, this $\nu
\rightarrow -\infty$ limit, would correspond to the singularity
$R\rightarrow 0$.  However, these equations do not explicitly show the actual evolution of $e^\lambda$ and $e^\nu$
as a function of $R$ and the $T\rightarrow \infty$ limit covers both the
$R\rightarrow R_{gb}$ limit as well as the further $R\rightarrow 0$ limit.
But did these equations really manage to show the progress of the
collapse to $R_b< R_{gb}$ region? Before we investigate this, note that,
 the solution for $e^\nu$ is of a highly discontinuous nature
By the phrase ``discontinuous nature'' we mean here the following thing:
As the central singularity is approached, $R\rightarrow 0$, the metric
coefficients for all the
regions of the fluid, whether internal ($r <r_b$) or on the boundary
($r=r_b$) should behave in a unique way. This obviously is not true for
$e^\nu$ in Eq. (8.20).

This hints that there is some tacit assumption which is not realized in
Nature (GTR) or there is a basic fault in the formulation of the problem.

Yet, one may try to ignore this suggestion because of our presumption
that ``trapped surfaces'' and ``event horizons'' are most natural
consequences of GTR collapse and on the specific plea that eventually $e^\nu
\rightarrow 0$ in the limit $T \rightarrow \infty$ for any $r$ inspite of
the discontinuous nature of the solutions.

However, such a plea would fail
for $e^\lambda$ where,  although for the boundary region, one finds
$e^\lambda$ to blow up,
$ e^\lambda \rightarrow \infty$, as it must, the value of $e^\lambda$
{\bf
 remains  finite} for internal points even when one is
supposed to approach the singularity $R(r, T) \rightarrow 0$. This is a
definite signature that {\bf there is a severe problem in the foundation of this
problem}. It is
surprising that neither the referee of the OS-paper not thousands of
 research workers who always claim that OS paper has categorically shown
the formation of BHs have ever noted this point!

\subsection{True Solution of the O-S Problem}

 For continued collapse, first, $R$ is supposed to touch the horizon $R
\rightarrow R_{g}$ and then dip below it to hurtle down:
the outermost surface gets trapped and a real event horizon
is formed, and then, even the outermost surface collapses to the central
singularity $R\rightarrow 0$.

In our attempt for a possible resolution of this physical anomaly with
regard to the discontinuous OS solutions, we see
from Eq. (8.9), that,
$T\rightarrow \infty$ if either or both of the two
following conditions are satisfied:
\begin{equation}
R_{gb} \rightarrow 0; \qquad T\rightarrow \infty
\end{equation}
and
\begin{equation}
y \rightarrow 1 ; \qquad T\rightarrow \infty
\end{equation}
OS {\em implicitly assumed} that $R_{gb}$ is {\em finite}
 and then $y \rightarrow 1$ and
then $y <1$ as $T\rightarrow \infty$

``For $\lambda$ tends to a finite limit for $R \le R_{gb}$ as $T$
approaches infinity, and for $R_b=R_{gb}$ tends to infinity. Also for $R \le
R_{gb}$, $\nu$ tends to to minus infinity.''
But,
while doing so, {\bf they completely overlooked
 the most important feature} of Eq. (8.9) (their Eq. 36),
that in view of the presence of  the $T \sim \ln {y^{1/2} +1\over y^{1/2} -1}$ term,
 in order that $T$ {\bf is definable at all}, one must have
\begin{equation}
y \ge 1
\end{equation}
For an insight into the problem, we first focus attention on the outermost
layer where $y_b = R_b/R_{gb}$, so that the Eq. (8.22) becomes
\begin {equation}
R \rightarrow R_{gb}
\end{equation}
But the condition (8.23) never allows $R$ to plunge
below $R_{gb}$:
\begin{equation}
R \ge R_{gb}
\end{equation}
Thus a careful analysis of the GTR homogeneous dust problem as enunciated
by OS themselves actually tell that trapped surfaces can not be formed
even though one is free to chase the limit $R\rightarrow R_{gb}$.

And, it can be verified that the OS paper, actually, only studied the approach to $R \rightarrow
R_{gb}$ limit for the outer boundary $r=r_b$ and
 the corresponding evolution of the internal regions $r < r_b$,
without ever showing $2GM (r, T) \le R$ at any $r$.
{\em Therefore, this work never really showed the formation of trapped
surface}. Most likely, the fact that OS studied the
the tendency for the formation of the ``horizon'' even in the context of
an actual internal soltution, is mistaken as the evidence for the
formation of a ``trapped surface''.

Yet, it is,
 indeed possible to attain the limit $R\rightarrow 0$ as $T\rightarrow \infty$ by
 by envisaging
\begin{equation}
R_b \rightarrow R_{gb} \rightarrow 0; \qquad {R_b\over R_{gb}} \ge 1; \qquad {2G
M_b\over R_b} \le 1
\end{equation}
This means that, the final gravitational mass of the configuration is
\begin{equation}
M_f (R=0) = M_g (R_b=R_{gb}) =0
\end{equation}
But then, for a dust or any adiabatically evolving fluid
\begin{equation}
M_i = M_f = constant
\end{equation}
Therefore, we must have $M_i = 0$ too. And {\em for a finite value of}
$R$, this is possible only if $\rho =0$. But for a dust $\rho =\rho_0 =
mn$ and therefore, we have $n=0$. Finally, the total number of the baryons in the configuration
\begin{equation}
N=0
\end{equation}
From, a purely mathematical view point,
the $N=0$ limit can be described as
\begin{equation}
r =r_b \rightarrow 0; \qquad r/r_b \rightarrow 1
\end{equation}

Although, we took pains to arrive at this above inherent structure of the
O-S dust, it could have been obtained much earlier, in a direct fashion,
simply from our Eq. (8.10) defining $y$ (Eq. [32] in the O-S paper).

First note that in Eq. (8.9), there is a basic constraint on the nature of
$y$, which is more elementary that what we have already pointed out : $y
\ge 1$. This most elementary constraint is simply that $y$ {\em must be not be negative}:
\begin{equation}
y \ge 0
\end{equation}
Remember, if we really assume $R_{gb} \neq 0$ the second term $r_b
R/R_{gb} r$ of  Eq.(8.10)  can be made arbitrarily small as the collapse
proceeeds $R\rightarrow 0$.  Remember here that $r$ and $r_b$ are comoving
coordinates and are fixed by definition. So for any interior region $r$
seperated from the boundary by a finite amount $r <r_b$, $y$ {\bf becomes negative}
in contravention of Eq.(8.31) if $R_{gb} \neq 0$! This is alleviated if
either or both of the two following conditions are satisfied : (i) $r=r_b$, as
derived above or (ii) $R_{gb} =0$, which again leads to the previous condition.

{\em Thus had O-S carefully noted this simple point, they would probably
not have
proceeded with the rest part of their paper which hints at the
formation of a finite mass BH in a completely erroneous manner}. But, it
is much more surprising that thousands of research workers claiming for
the theoretical evidence for existence of BHs never had the time to
actually carefully read this paper.

So, mathematically, the Eqs. (8.17) and (8.18), in a self-consistent manner
 degenerate to  defininite limiting forms:
\begin{equation}
e^{-\lambda} \approx 1 - \left(e^{-T/R_{gb}} +1\right) \rightarrow 0
\end{equation}
or,
\begin{equation}
e^\lambda \rightarrow \infty
\end{equation}
\begin{equation}
e^\nu \approx e^{\lambda - 2T/R_{gb}} \left\{e^{-T/R_{gb}} +1\right\} \rightarrow 0
\end{equation}
Thus, technically, the final solutions of OS are correct, except for the
fact {\bf they did not organically incorporate neither the crucial} $y \ge
0$ nor the $y >1$ conditions in
the collapse equations. And all we have done here is to rectify this
colossal  lacunae to fix the value of $R_{gb}=0$.

And obviously $\tau \propto R_{gb}^{-1/2} \rightarrow \infty$ just like $T$ as $R
\rightarrow R_{gb} \rightarrow 0$. Very strictly, since $r_b =0$, for $N=0$,
this above equations point to an inherent faulty formulation of the
problem in terms of a strict $p=0$ EOS.
Although, OS did not find a more explicit
expression for proper time for collapse, for a homogeneous dust,
it is possible to find
explicity an expression for $\tau_c$
 for collapse upto the central singularity, in a manner analogous to the
Newtonian solution of Sec. 1.
provided one assumes that the
{\em collapsing dust was at rest} at $\tau =0$ at $r_b=R=R_\infty$ [9, 48]:
\begin{equation}
\tau_c = {\pi \over 2} \left({ 3\over 8\pi G \rho(0)}\right)^{1/2}
\end{equation}
where $\rho(0)$ is the density of the dust {\em when it was at rest}; $U(0)=0$.
This equation also follows from Eq. (8.14).
But a dust can never be at rest: ``When the pressure vanishes there are no
solutions to the field equations except when all components of $T^i_k$
vanish. With $p=0$ we have the free gravitational collapse of matter''
[26]. Therefore, from dynamics, a $p=0$ EOS necessarily gives $\rho =0$ in
hydrostatic equilibrium and this can be verified in a straight forward
manner from the Oppenheimer- Volkoff [25] equation:
\begin{equation}
{dp\over dR}=- {p+\rho(0) \over R(R-2GM)}(4\pi p R^3 + 2G M)
\end{equation}
This means that $\rho =0$, a conclusion, already arrived from a different
consideration. From the view point of thermodynamics, any physically
meaningful EOS will yield $\rho =0$ if $p=0$ irrespective of whether the
fluid is in hydrostatic equilibrium or not. Now, more precisely, we see that, $\tau_c \rightarrow
\infty$ just like $T\rightarrow \infty$. And also, for a homogeneous dust
$\rho =0$ means that the $M= (4\pi/3) R^3 \rho =0$ if $R$ is finite. On
the other hand, the mathematical expression for the fact that ``a dust can
never be at rest'' would mean that, technically, $R_\infty =\infty$ and
the dust solution can not be matched with the collapse of any star of
finite radius. If $R_\infty =\infty$, the mass of the star is $M=\infty$
unless $\rho =0$. The $M=\infty$ condition, again, can not be matched with
any star of finite mass. In any case, if $R_\infty=\infty$, the proper time taken
to attain any finite value of $R$ would be $\tau =\infty$ because $v= finite$.

This shows that the problem of ``collapse of a homogeneous dust'' is
fictitious and can not be formulated in a physically meaningful way. We
have noted in Sec. 2 that this is true for Newtonian dust collapse problem
too. And
finally note that the expression for $\tau_c$ is exactly the same in both
Newtonian physics and GTR. Why it happened like this? f This question has
always been {\bf avoided in the past by assuming that this was just a matter of
coincidence}. Actually it is not so. As we emphasized, {\em the dust solutions
are strictly valid, both in the Newtonian case and GTR case, in the limit
of $\rho \rightarrow 0$. And in this limit GTR merges with Newtonian gravitation}.

\subsection {Inhomogeneous Dust}
It might appear that, this fictitious nature of the GTR dust collapse problem
explained above resulted from  the  assumption of homogeneity and
inhomogeneous dust solutions may be closer to physical reality.
Inhomogeneous dust solutions do not admit any exact solution and depending
on the various subtle approximations used, it is indeed possible to have
wide variety of solutions. In particular, several authors have attempted
to point out that it is not necessary that a BH is produced. On the other
hand, the resultant singularity could be a ``naked'' one, a singularity
not clothed by an event horizon, and hence may be visible to an outside observer.
This would be in contrary to the ``cosmic censorship conjecture'' of
Penrose [42], for which, there is no general proof, and several authors claim
that there could be counter examples. The point is that there is no
analytical solution for the collapse problem for a physical fluid or even
for an inhomogeneous dust. And though, the famous singularity theorems
[16, 41, 49, 50]  (apparently) confirm that collapse to a singular state
is inevitable not only in spherical symmetry but even when there is
deviation from it, the highly idealized (apparent) example of OS collapse
can not be generalized to predict the formation of BHs. However, in a most
general fashion, by putting the work of Tolman [39] in proper physical
perspective, we show below that, even for an inhomogeneous dust, in order
to approach the $R\rightarrow 0$ limit, we must have $M=0$.

By recalling that for a dust
\begin{equation}
{\dot R} = {dR\over d\tau} =U$; \qquad  $\Gamma= {dR\over \sqrt {-g_{rr}} dr}
= {R'\over B}
\end{equation}
we can reframe the results of Tolman [39] in our notation as
\begin{equation}
U^2 = \Gamma^2 -1 + {2 G M(r)\over R }
\end{equation}
\begin{equation}
F(R) = 2 G M(r)
\end{equation}
\begin{equation}
1 + f(r) = \Gamma^2 >0
\end{equation}
\begin{equation}
U= \Gamma v
\end{equation}
These equations explicitly show that $\Gamma = \sqrt {1+f}$ and $M(r)$ are
independent of $\tau$ and hence are  constant at a fixed comoving
coordinate $r=r$. Recall that this result was already obtained in Sec. 7.
As before, $\Gamma \rho$ represents the energy
density measured by $S_\infty$ and hence $\Gamma$ is necessarily finite.
By POE and STR, $v \le 1$ is finite, and therefore Eq. (8.41) tells that $U$ is
finite too. More explicitly, the above equations yield:
\begin{equation}
v^2 = {U^2 \over \Gamma^2} = {\Gamma^2 -1 + {2GM \over R}\over \Gamma^2}
\end{equation}
In order that, $v^2\le 1$ is bounded, the foregoing Eq. demands that
\begin{equation}
{2GM\over R} \le 1
\end{equation}
which is essentially the same constraint $y_b \ge 1$ obtained earlier for
homogeneous dust.
This explicitly shows that trapped surfaces are not formed for dust collapse.
Assumption of positivity of mass, now obviously yields
\begin{equation}
M_f \rightarrow 0; \qquad R\rightarrow 0
\end{equation}
But, for a dust, $M \propto F(r)$ is independent of $t$ and hence is constant.
Thus, if an inhomogeneous dust is to collapse to $R=0$,
 we have $M=0$ even when $R\neq 0$. This means that for
all dust, which might be envisaged to collapse to $R=0$, we have
\begin{equation}
{2 G M\over R} =0; \qquad if,~   R\neq 0
\end{equation}
Then Eq. (8.42) shows that, for a value of $R\neq 0$, we must have
\begin{equation}
v^2 = {U^2 \over \Gamma^2} = {\Gamma^2 -1 \over \Gamma^2} = {\rm constant}
\end{equation}
In spherical geometry and in the presence of gravity, the foregoing
condition can be satisfied only if there is no dust collapse at all:
\begin{equation}
v \equiv 0; \qquad \Gamma_{dust} \equiv 1
\end{equation}
Why must the value of $v$ in the GTR dust collapse be zero? Unlike a
physical fluid, a dust is really not a fluid because there is no
mutual interaction between the particles; it is a mere incoherent
collection of particles. A physical expression for this statement is that
the sound speed in dust $\equiv 0$ (unless $\rho =0$). Therefore, unlike the case of a physical fluid,
one should be able to analyze the spherical dust collapse problem as a gross
addition of $N$ incoherent processes. And in either case, {\em one must obtain the
same result}. But, as soon as we analyze the problem of motion of a single
dust particle in the gravitational field of the underlying dust, the
problem becomes a two-body problem. As emphasized before, unlike in
Newtonian gravity, in GTR, a two-body problem is necessarily
non-stationary and should result in the emission of gravitational waves.
But, when treated as the real spherical case, the field is static and
there should not be any gravitational radiation. This dichotomy can be
resolved only if $v =0$ !

\subsection {Proper Time for General Dust Collase}
Since $U=dR/d\tau$, the proper time for collapse, as obtained by Tolman [39] was
\begin{equation}
\tau=\int {dR\over \sqrt{ f+ {F\over R}}}
\end{equation}
In our notations this equation looks like
\begin{equation}
\tau=\int {dR\over \sqrt{ \Gamma^2 -1 + {2GM \over R}}}
\end{equation}
Upon integration, we have:
\begin{equation}
\tau = {2 \over \sqrt{1 -\Gamma^2}} \tan^{-1} \sqrt{{\Gamma^2-1 +
{2GM\over R} \over 1-\Gamma^2}}
\end{equation}
Since, $\Gamma^2 =1$ for a dust, we find that, the proper time taken to
arrive at any finite value of $R$ (not only $R_{gb}$ or $R=0$) is $\tau
=\infty$. This result is in accord with the essential hypothetical nature
of the formulation of the problem.
We have already shown that  for a dust with $\rho =\rho_0$,
the condition $M =0$ implies $N =0$. Taken together, the above two results
express the fact that any macroscopic collection of matter (or energy) is
necessarily characterized by a finite, howsoever small, pressure, and,
dust solutions can never represent the actual collapse of a physical
fluid. And in any case, we explicitly showed that the interpretation of OS
that, dust collapse results in the formation of an event horizon was
incorrect, and was obtained by overlooking the fundamental message of the collapse
equations that we must have $ y >1$.
\section{Collapse of Physical Fluid}
It might appear that the explicit result shown above that $M_f \rightarrow 0$ for continued
collapse $R\rightarrow 0$ could be derived for a dust because of the
inherent simplicity caused by the time independence of $\Gamma $. Before we
investigate this aspect, we would emphasize that, for studying the
collapse of a physical fluid, it is absolutely necessary to incorporate
the radiation transport aspect in an organic fashion. For a physical
fluid, $\Gamma$ will, in general, be time dependent and no firm conclusion
about the evolution of the fluid may  be drawn. Yet one must be able to
formulate the problem properly even if it may not be solved. The collapse
equations were generalized to incorporate the presence of radiation by
several authors [11, 35, 44, 45, 55].

The radiation transport may be handled in two limits:
for small opacities, one may use the geometrical optics form of the
radiation part of the stress tensor
\begin{equation}
E^{ik}= q k^ik^j
\end{equation}
where $q$ is both the energy density and the radiation flux in the proper
frame and $k^i=(1;1,0,0)$ is a null geodesic vector so that $k^ik_k=0$.
For very large opacities, one may adopt the diffusion approximation [9,45]. For
simplicity following Misner [11] and Vaidya [35], we will first treat the radiation in the
geometrical optics limit. Then, we would realize that our global constraint
equations are actually independent on the form of $E^{ik}$ (although other
equations may change) because the modified definition of $M$, in the
presence of radiation, absorbs all the radiation terms. The other
equations of course will change in keeping with the changing form $E^{ik}$
and we would nor require these equations.

All one has to do now is to repeat the exercises  for an adiabatic fluid
outlined in Sec.7 by replacing the pure matter part of energy momentum
tensor with the total one:
\begin{equation}
T^{ik} = (\rho +p) u^i u^k + p g^{ik} + q k^i k^k
\end{equation}
Then the new $T^0_0$ component of the field equation, upon integration,
yields the new mass function:
\begin{equation}
M(r,t) = \int^r_0 4 \pi R^2 dR ( \rho + q v +q) = \int^r_0 d{\cal V }
[\Gamma (\rho + q) +q U]
\end{equation}

Had we treated the radiation transport problem without assuming a
simplified form of $E_{ik}$ and, on the other hand, in a most general manner,
following Lindquist [45], we would have obtained:
\begin{equation}
M(r,t) =  \int^r_0 d{\cal V }
[\Gamma (\rho + J) +H U]
\end{equation}
where
\begin{equation}
J= E^{00} = E^{ik} u_i u_k =q = comoving~ energy~ density
\end{equation}
and
\begin{equation}
H = E^{0R} = average~ radial~ flux
\end{equation}
This definition of $M$ may be physically
interpreted in the following way: while
$(\rho+q)$ is the locally measured energy density of matter and radiation,
$\Gamma(\rho+q)$ is the same sensed by $S_\infty$ ($\Gamma \le 1$). Here,
 the radiation part may be also explained in terms of `` gravitational red -shift''.
 And the
term $HU$ may be interpreted as the Doppler shifted flux seen by $S_\infty$.

Although, the collapse equations, in general will change for such a general
treatment of radiation transport,
{\em  the generic constraint equation involving} $\Gamma$ {\em
incorporates  this
new definition of} $M$ and remain unchanged:
\begin{equation}
\Gamma^2 = 1 + U^2 - {2 G M\over R}
\end{equation}
In the following, we list
the other major collapse equations for the simplified form of $E^{ik}$ only:
\begin{equation}
D_{\rm r}M=4\pi R^2\left[\Gamma (\rho +q)+ U q\right]
\end{equation}
\begin{equation}
D_{\rm t}M =- 4 \pi R^2 p U- L (U+\Gamma)
\end{equation}
\begin{equation}
D_{\rm t}U=-{\Gamma \over \rho+p} \left({\partial p \over \partial
R}\right)_{\rm t} -{M+4\pi R^3 (p+q)\over R^2}
\end{equation}
\begin{equation}
D_{\rm t}\Gamma=-{U\Gamma\over \rho +p}\left({\partial p\over \partial
R}\right)_{\rm t} +{L\over R}
\end{equation}
where the comoving luminosity is
\begin{equation}
L=4\pi R^2 q
\end{equation}
First note that, with the inclusion of $q$ in this problem the qualitative
argument for the inevitability of the formation of a BH mentioned in the
context of adiabatic collapse [22] completely breaks down. At any rate,
these arguments only showed that the value of $\Gamma$ should steadily
decrease in a GTR collapse in agreement with our qualitative arguments of
Sec. 1 $(\Gamma_s \rightarrow 0)$.
\subsection {Singularity Theorems}
Even if there is no
question of a strict exact solution (numerical or analytical) for such a fluid,
it is believed by practically all the authors
that a physical fluid will necessarily collapse to a singularity in a
finite proper time; and the debate hinges on whether the singularity would
be a BH or a naked one. For a general GTR collapse, there are several notions
for the definition of a singularity. And although all such notions may not be
fully compatible to one another, the most fundamental definitions of
spacetime singularities involve incompleteness of {\bf timelike or null}
worldlines [9 16 49, 50]:

``The study of timelike curves is fundamental to the study of fluids in
GTR, for the world lines of fluid particles (the integral curves of the
fluid 4-velocity) form a family of timelike lines''[16].

``Despite these complications, which mean that it is difficult to define
what is meant by a singularity, consensus has been reached that a
sufficient criterion for the existence of a singularity is a proof that
there are incomplete timelike or null geodesic in an inextendible space-time''
[16].
The modern conviction in the inevitability of the occurrence of spacetime
singularities in a general gravitational collapse of a sufficiently massive
configuration stems on the strength of singularity theorems. Probably, the
first singularity theorem, in the context of spherical collapse, was
presented by Penrose [41] where it was explicitly shown that once a
trapped surface is formed, $2G M(r,t) /R >1$, the collapse to the central
singularity is unavoidable.
 Since then many authors like Hawking, Geroch, Ellis, including Penrose
himself, have proposed various forms of singularity theorems, and a
history of the evolution of this line of research may be found in the
review article [16].  The singularity theorems used generic topological
arguments based on physically reasonable and general conditions on the
spacetime structure  : ``A spacetime $M$  necessarily contains
incomplete, inextendable timelike or null geodesics if, in addition to the
Einstein's equations, the following four conditions hold''[9]

(1) $M$ contains no closed timelike curves demanding causality is not violated

(2) At each event in $M$ and for each unit timelike vector (like four
velocity) $u^i$, the stress energy tensor satisfies

\begin{equation}
\left(T_{ik} -{1\over 2} g_{ik}\right)u^{i} u^{k}\ge 0
\end{equation}
For a perfect fluid, this condition reduces to
\begin{equation}
\rho +3p \ge 0
\end{equation}
This condition is also called ``strong energy'' condition and, obviously,
very reasonably demands that the stresses remain positive or even if some
of them become negative, they remain sufficiently bounded.

(3) The ``manifold'' is general in the sense that every timelike or null
geodesic with unit vector $u^i$ passes through at least one event where
the curvature tensor is not lined up with it in a certain specific way.

(4) And finally the manifold should contain a trapped surface either in
the past or future

The major utility of the singularity theorems are not intended to be for
the spherical case, where, the formation of trapped
surfaces,  and singularities seem to be almost a foregone conclusion:
``Since horizons and accompanying  trapped surfaces are necessarily
produced by slightly nonspherical collapse and since they probably also
result from moderately deformed collapse such collapse presumably produces
singularities - or a violation of causality, which is also a rather
singular occurrence'' [9]. It is only for the highly nonspherical
configurations that assertion about the inevitability of the formation of
singularity appeared to  be difficult, and the singularity theorems are
particularly relevant. Despite such a theoretical stand,  the fact is
that, the {\em formation of a trapped surface remains an assumtion} even
for the simplest spherical case.

It is clear therefore, that, {\em if we are able to show that trapped surfaces
are not formed even for the most idealized case of a nonrotating perfectly
spherical perfect fluid not having any resistive agent like a strong
magnetic field, certainly trapped surfaces would not form in more
complicated situations}.

\subsection {Final Proof}
We have already shown explicitly that for a dust collapse $y \ge 1$ implying
that trapped surfaces are not formed. For a dust $\Gamma$ was constant. In a
general case $\Gamma$ is not so, yet, we can easily find a global property
of the GTR collapse problem which is absolutely independent of the actual EOS or any
other details. Although we tried to extend the exiting general framework
for handling the Einstein equations for greater physical insight it was
really not necessay to introduce the physical velocity   and
to find the relationship between two global quantities $U$ and $\Gamma$.
Unfortunately, moxt of the texts on GTR do not give explicit discussion on
the physical velocity, and consequently many readers/referees may be
confused about the actual expression for $v$ in GTR. Thus in the
following, we would first derive our central result without explicitly
introducing any $v$ at all.

\subsubsection{Proof Without Physical Velocity}
 Whenever we say that we are studying the collapse problem,
 by definition, we are studying radial timelike or null worldlines of the
material particles of the fluid or the embedded radiation.
with metrics : $ds^2 \ge 0$ (if
the signature of the metric is 1, -1, -1, -1).

Let us consider  comoving coordinates, attached to particles.
  In
spherical symmetry, the (comoving) radial coordinate is most appropriately
defined by a marker $r$ inclosing fixed number of baryons - by definition,
there is no question of any coordinate singularity here, i.e, $g_{00} \ge
0$.  For purely radial motions, one may ignore the
angular part of the metric to write:
\begin{equation}
ds^2 = g_{00} dt^2 + g_{rr} dr^2
\end{equation}
Again, {\em by definition}, the worldlines of photons or material
particles are null or time like, i.e., $ds^2 \ge 0$, so that
\begin{equation}
g_{00}\left[1 +\left( g_{rr} dr^2\over g_{00} dt^2\right)\right] \ge 0
\end{equation}

We have found in Section 7 and 8 that, there exists a positive
definite quantity
\begin{equation}
\Gamma^2 \equiv {1\over -g_{rr}} \left({dR\over dr}\right)^2=
 1 + {1\over g_{00}}\left({dR\over dt}\right)^2 - {2 G M(R) \over R c^2} \ge 0,
\end{equation}
which is not negative because $g_{rr}$ is negative in the given basis.
By transposing the foregoing Eq., it follows that
\begin{equation}
\Gamma^2 \left[1 +\left( g_{rr} dr^2\over g_{00} dt^2\right)\right] = 1 -
{2 G M(r,t) \over Rc^2 }
\end{equation}
In Eq.(9.16), $g_{00}$ is positive so that
\begin{equation}
\left[1 +\left( g_{rr} dr^2\over g_{00} dt^2\right)\right] \ge 0
\end{equation}
And since $\Gamma^2$ is positive  the
L.H. S. of Eq.(9.18) is positive. And so must be its R.H.S. Then Eq. (9.18) yields:
\begin{equation}
 1 - {2 G M(r,t) \over Rc^2 } \ge 0
\end{equation}
Although, {\em comoving coordinates}, by definition, do not involve any
singularity unlike external Schwarzschild coordinates, in a desperate
attempt to ignore this foregoing small derivation, some readers might
insist that there could be a coordinate singularity somewhere so that
 $g_{00}$ could be negative in Eq.(9.16). Even if one accepts
this incorrect possibility for a moment, our eventual result survives such
incorrect thinking in the following way.

If $g_{00}$ were negative in Eq.(9.16), we would have
\begin{equation}
\left[1 +\left( g_{rr} dr^2\over g_{00} dt^2\right)\right] < 0
\end{equation}
But the determinant of the metric, $g = R^4 \sin^2 \theta ~g_{00} ~g_{rr}$
is always negative (Landau \& Lifshitz [30]) so that, in this wild
situation, we would have $g_{rr} >0$. Then it follows from Eq.(9.17) that
$\Gamma^2 <0$ so that the L.H.S. of Eq.(9.18) is again positive. Hence the
R.H.S. of Eq.(9.18) too must be positive, and we get back Eq.(9.20) whence
it follows, {\em in a most general fashion}, that
\begin{equation}
{2 G M(r,t) \over Rc^2 } \le 1
\end{equation}
This shows that {\bf trapped surfaces do not form}, and further, by
invoking the Positive Mass theorems, it follows that
\begin{equation}
M(r,t) \rightarrow 0; \qquad ~ as~ R\rightarrow 0
\end{equation}
Note that {\bf this derivation could be achieved without explicitly
introducing} $v$ ! So even if our interpretation of $v$ were incorrect,
the basic result {\bf remains unchaged}.

\subsubsection{Proof By Introducing v}
The above derivation can of course be made in an elegant fashion by
explicitly invoking the concept of a physical 3-velocity.
 We simply substitute the global relation
\begin{equation}
U=\Gamma v
\end{equation}
into the right hand side of another global constraint
\begin{equation}
\Gamma^2 = 1 + U^2 - {2G M(r,t)\over R(r, t)}
\end{equation}
to obtain
\begin{equation}
\Gamma^2 = 1 + \Gamma^2 v^2 - {2G M(r,t)\over R(r, t)}
\end{equation}
Now by transposing, we obtain
\begin{equation}
\Gamma^2 (1 -v^2) = 1- {2G M(r,t)\over R(r, t)}
\end{equation}
It may be rewritten as
\begin{equation}
{\Gamma^2\over \gamma^2} = 1-{2G M(r,t)\over R(r, t)}
\end{equation}
The foregoing beautiful equation may be termed as the ``master equation'' for spherical
gravitational evolution of a system of a fixed number of baryons.
Since the left hand side of the two foregoing equations are $\ge 0$, so
will be their right hand side:
\begin{equation}
1-{2G M(r,t)\over R(r, t)} \ge 0
\end{equation}
Thus we obtain the most fundanental constraint for the GTR collapse (or
expansion) problem, in an unbelieveably simple manner, as
\begin{equation}
{2G M(r,t)\over R(r, t)} \le 1 ; \qquad {R_{g} \over R} \le 1
\end{equation}
This is the ultimate proof that {\em trapped surfaces are not allowed by GTR}.
However, we foresee that, in view of the astonishing simplicity involved
in this proof, many
readers would find it difficult to appreciate it.
\subsection{Positive Energy Theorems}
Unlike Newtonian physics, there is no clear notion about what may be
precisely called ``energy'' in GTR. For example in a non-stationary
problem, like the two-body problem alluded to before one can not at all
precisely define any energy for the system. However, energy can be
 defined for a static and stationary gravitational field.
And  for a static system, such as one we are discussing, there exists a
well defined notion of ``global energy''
 with respect to $S_\infty$.  Although, a global energy of an
isolated system can be defined, it can not be asserted before hand whether
this would be positive, zero or negative.  And it is practically a branch
of research in gravity theories to establish that the energy of an isolated
body is indeed non-negative  [15, 19, 43]. And it is believed that the
energy of an isolated body can not be negative. From physical view point, a negative value of
$M_b$ could imply repulsive gravity and hence not acceptable.

When we accept this theorem(s), we find that the fundamental constraint
demands that {\em if the collapse happens to proceed upto} $R\rightarrow
0$, i.e., upto the central singularity, we must have
\begin{equation}
M(r, t) \rightarrow 0; \qquad R\rightarrow 0
\end{equation}
Remember here that the quantity  $M_0 =m N$ (which is the baryonic mass of
the star, if there are no antibaryons) is conserved as $M_f \rightarrow 0$.
 Physically, the $M=0$ state may result when the
{\em negative gravitational energy} exactly cancels the internal energy, the
{\em baryonic mass energy} $M_0$ and any other energy, and  which
is possible in the limit $\rho \rightarrow \infty$ and $p\rightarrow
\infty$.

\subsection{Previous Hints}
While considering, the purely static GTR equilibrium configurations of dust,
Harrison et al. [6]
discussed long ago that spherical gravitational collapse should
come to a {\em decisive end} with $M_f=M^*=0$, and, in fact, this
understanding was formulated as a ``Theorem''

``THEOREM 23. Provided that matter does not undergo collapse at the
microscopic level at any stage of compression, then, -regardless of all
features of the equation of state - there exists for each fixed number of
baryons A a ``gravitationally collapsed configuration'', in which the
mass-energy $M^*$ as sensed externally is zero.'' (See Appendix 4)
Ironically, one of the co-authors of the above statement, later coined the word
``Black Hole'' [21].

In a somewhat more realistic way   Zeldovich and Novikov [56] discussed the
possibility of having an ultracompact configuration of degenerate fermions
obeying the EOS $p=e/3$ with $M\rightarrow 0$
and mentioned  the possibility of {\em having a machine for
which} $Q\rightarrow M_i c^2$. (See Appendix 5).

  It is widely believed that Chandrasekhar's
discovery that White Dwarfs (WD) can have a maximum mass set the stage for
having a gravitational singular state with finite mass. The hydrostatic
equilibrium of WDs can be approximately described by Newtonian
polytropes [46] for which one has $R\propto \rho_c^{(1-n)/2n}$, where
$\rho_c$ is the central density of the polytrope having an index $n$.  It
shows that, for a singular state i.e., for $\rho_c\rightarrow \infty$, one
must have $R\rightarrow 0$ for $n>1$; and Chandrasekhar's limiting WD
indeed has a {\em zero radius} [46]. On the other hand, the mass of the
configuration $M\propto \rho_c^{(3-n)/2n}$.  And unless $n=3$,
$M\rightarrow \infty$ for the singular state. One obtains such a result
for Newtonian polytropes because they are really not meant to handle real
gravitational singularities. Fortunately, in the low density regime, when
the baryons are nonrelativistic and only electrons are ultrarelativistic,
the EOS is $p\rightarrow e/3$ and the
corresponding $n\rightarrow 3$. Then one obtains a finite value of $M_{ch}$
- the Chandrasekhar mass.

Now when we apply theory of polytropes for
a case where the pressure is supplied by the baryons and not only by
electrons, we must consider GTR polytropes of Tooper [40]. It can be easily
verified from Eq.  (2.24) of this paper [40] that in the limit $\rho_c
\rightarrow \infty$, the scale size of GTR polytropes $A^{-1} \rightarrow 0$.
Further Eqs. (2.15) and (4.7) of the same paper  [40] tell that $M \propto K^{n/2}
\propto \rho_c^{-1/2} \rightarrow 0$ for $\rho_c\rightarrow \infty$. Thus,
a {\em proper GTR extension of Chandrasekhar's work would not lead to a BH
of finite mass}, but, on the other hand, to a singular state with
$M\rightarrow 0$.

In a different context, it has been argued that {\em naked singularities}
produced in spherical collapse must have $M_f=0$ [27].

\section{Probable Regimes of Confusion}
Although, the Positive Energy Theorems probe whether $M_b$ can not only be
zero but even negative, and although many of the so-called naked
singularity solutions coorespond to zero gravitational mass [27], we can
foresee that many readers would have difficulty in accepting our result.
It may be so, because our result not only shows that $M_f=0$ (for continued
collapse only), but it also explicitly invalidates
(i) One of the greatest plinths of modern gravity research, namely, the
singularity theorems. (ii) Further this work effectively removes one of
the most cherished mystique of not only modern science but also of the science
of Newtonian era, namely, the existence of (finite mass) Black Holes. And
naturally, almost instinctively, there may be a tendency to reject this work
on the basis of tangential and vague reasons. And though, we have taken
great care in developing several ideas which, normally not only experts but all
serious students of GTR are expected to know, in the face   of the likely
strong revulsion, several genuine or apparent confusions
may creep up. In fact an abridged version of the present work [2] was rejected
by the Physical Review Letters (PRL) solely on the basis of such confusions:
\subsection{Baryonic and Gravitational Mass}
One of the Divisional Associate Editors of PRL rejected this paper on the
basis  that a $M=0$ state corresponds to
zero baryon number $N=0$, and unless either all the baryons can physically
fly away or the agents of the radiation mechanism like
leptons (like neutrinos) or photons carry baryon number in a bizarre
physics, our result is not acceptable! Very clearly, despite formal
knowledge of GTR, the reader, because of instinctive Newtonian notion, equated
the gravitational mass with the baryonic mass: $ M \equiv M_0 = m N$
(incorrect), and hence equated a $M=0$ scenario with $N=0$ one.

Although already emphasized, the gravitational mass of an isolated system
is just the aggregate of all kinds of energy associated with it, and for
any bound system, necessarily $M < M_0$. This is something like the fact
that the mass of an atom or a nucleus is always less than the aggregate of
the masses of the individual constituents, like electrons, protons or neutrons:
\begin{equation}
M = M_0 + E_g + E_{in} + E_{kinetic}
\end{equation}
Here the gravitational energy term is always negative (even if $ M <0$)
and non linear. In the weak gravity regime it is $\sim -G M^2 /R$, while
its GTR form is given by Eq. (1.21). As collapse proceeds, the grip of gravity
becomes tighter, and this is effected by the non linear nature of $E_g$
As a result, the value of $M$, in general, steadily decreases in any
gravitational collapse.  If the collapse can be halted at an intermediate
state, obviously, the value of $M< M_i$ while $M_0$ remains unchanged.
At any intermediate state, the value of $\mid E_g\mid$ and $E_{in} + M_0$ would
be closer than their difference in earlier epochs although individually,
the value of $\mid E_g\mid$ and $E_{in}$ steadily increases. Then, it is a
natural consequence that if we have a continued collapse, the value of
$M$  will hurtle downward and the system would try to seek a state of
``lowest energy''. In GTR, i.e., in Nature, the lowest energy corresponds
to $M=0$ and not to its Newtonian counterpart $E_N =0$ (incorrect). Thus,
 If we remove the possibility of the occurrence of a repulsive gravity (
negative $M$), then the bottom of the pit would be at $M_f =0$. At this
state, both $ E_g$ and $E_{in}$ would be infinite but  of opposite sign
and separated by a finite gap $M_0$ much like what happens in a
renormalized Quantum Field Theory.

There could be another confusion here as to how can $\mid E_g\mid$ be
infinite when $R =0$. This depends on how fast the value of $M \rightarrow
0$ with respect to $R\rightarrow 0$ and is perfectly allowed for a
singular state.

\subsection {Principle of Equivalence}
Even though there are many published results suggesting $M=0$ in
connection with naked singularities, our work might be singled out with
the plea that a $M=0$ result violates POE. We repeat once again that, POE
only says that the local nongravitational laws of physics are the same as
the corresponding laws in STR. For example, this would mean that the
Stephan- Boltzman law which tells that the emissivity of a black body surface is
$\propto T^4$, remains unchanged. POE does not impose any limit on the
value of $T$ itself and hence on the total amount of radiation emitted from the black
body surface. POE does not say that only a certain percentage of the
initial total mass energy $M_i$ can be radiated in the process, POE has
got nothing to do with either the imposition of any  additional local
constraint (such as a maximum value of $T$) or any global issues.

If one would invoke POE to debar phenomenon which are not understandable
in Newtonian notions (like $M \equiv M_0$, incorrectly) GTR itself is to
be discarded. With such a viewpoint,  all work on Positive Energy Theorems
are to be considered as  redundant and unnecessary because in STR, the
mass-energy of a system which was positive to start with can never be negative.

\subsection {Matter - Antimatter Annihilation ?}
In STR, there is no gravity and hence there is no Kelvin- Helhmoltz
process, neither could there be any real finite material body held
together by any long range force ( a plasma has to be confined by external
electromagnetic fields). And there could be a naive
 idea that the entire initial mass energy
may be radiated only if there are processes like $e^+ e^- \rightarrow  2\gamma$. If
this is envisaged as the only way to generate radiation (in this case
photons), it must be remembered that such a thing refers to systems having
total lepton mummer or total baryon number as zero. For matter consisting
of a definite baryon number and lepton number there can not be any energy
extraction by this process. Yet such matter radiates because of normal
electromagnetic processes like Bremmstralung or Compton processes, or by
nuclear processes like $ p+p \rightarrow \pi^0 \rightarrow 2\gamma$. Actually at
very high densities and temperatures, in astrophysical scenarios, energy
is liberated by the so-called URCA or weak interaction processes involving
emission of $\nu \bar \nu$. Whatever be the process, if the global Kelvin-
Helmotlz process heats up the matter to sufficiently high temperature near the
singularity (to which everybody agrees), the center of mass energy of the
colliding particles, like, electrons,
protons, neutrons, quarks or whatever it may be, will be accordingly high enough.
 And
in this limit, for an individual collision, the colliding particles
 can radiate not only an
energy equal to their rest mass but any amount higher than this. The
easiest example would be that an $e^- - e^+$ collider can generate
particles (photons, neutrinos, quarks etc.) much heavier than $0.5 MeV$. And it should be also remembered
that when we say that the entire $M_i c^2$ may be radiated, we do not mean
that this happens in a flash as is the case for matter-antimatter annihilation.
On the other hand, in gravitational collapse, it is the integrated radiation
over the entire history of the process we are concerned with.

\subsection {Confusion Between $\Gamma$ and $\gamma$}
Although, this point, too has been already discussed several times in this
work, there is some chance of a genuine confusion because of similar confusing
references about the nature of $U$ in the published literature. May and
White [32] correctly described $U$ as the ``1-component of the 4-velocity in
a Schwarzschild coordinate system''.
The
radial component of the 4-momentum in the Schwarzschild coordinate is
\begin{equation}
p^R = m {d R \over d\tau} = m U = m \Gamma v
\end{equation}
Note the notion of $\Gamma$ a global one and is not defined in terms of local quantities unlike the
Lorentz factor $\gamma$. In the context of the exterior SM, we have
specifically seen that $\Gamma ={\tilde E} /m$ is the energy per unit mass
as seen by $S_\infty$. This energy is smaller by a factor $(1- R_{gb}/R)^{1/2}$
than the local energy per unit mass, i,e, $\gamma$. The physically valid
initial condition, that in the absence of external fields, the test
particle at $R=\infty$ must be at rest in the absence of gravity ensures
that $\Gamma \le 1$.

However,  Hernandez and Misner [55], somewhat,
confusingly wrote  $U$ ``is therefore some sort of fluid 4-velocity or
momentum per unit mass'',
and this statement may be misinterpreted to imply that
$U$ is the ``momentum per unit mass'' in the STR sense.
 The actual (STR) radial component of fluid 4-velocity is to be defined with
respect to the LIF, and, we have repeatedly mentioned that, the same is
\begin{equation}
u^{\hat R} = \gamma v
\end{equation}
where, obviously, $\gamma \ge 1$. And the corresponding 4-momentum
component is
\begin{equation}
p^{\hat R} = \gamma m v
\end{equation}
The ``momentum per unit mass'' (in the sense of STR) is therefore $= \gamma v$. In fact the
abridged version of this work [2, 3]
 was originally rejected by PRL on the basis of this confusion!

\subsection {Spacelike Worldline ?}
In Section 6, we have discussed that, if one presumes the existence of a BH
of finite mass and a finite $R_{gb}$, and studies the motion of a test
particle (in a vacuum) the value of $v_{ex} \ge 1$ for $R \le R_{gb}$. And,
for a finite $R_{gb}$, this happens because the static external SM has a
coordinate singularity at $R_b=R_{gb}$ and therefore, one must restrict its
validity to $R > R_{gb}$. On the other hand, one must look for an appropriate
coordinate system which is of dynamic nature, like a COF or DSF, which are
bound to be singularity free if we have formulated the problem correctly.
This matter should have been rested at this point because the actual worldlines are by definition
timelike or null. In particular, when one studies the evolution of an
element of fluid and not of a test particle in vacuum, one can always
define a singularity free COF or DSF. Therefore, the entire discussion on
External {\em vacuum} SM or {\em vacuum} Lemaitre coordinate or {\em
vacuum} Kruskal coordinate have no direct
relevance for the collapse problem. To quote Weinberg, ``this discussion on Schwarzschild
singularity does not apply to any gravitational field actually known to
exist anywhere in the universe. Indeed, it does not even apply to
gravitational collapse'' [48].

Still we painstakingly discussed these issues so that the reader can
appreciate our work in a broader perspective. And, yet,
some reader may desperately try to reject our work by imagining that we
were working with the External SM, and the $v^2$ involved in Eq. (9.18)
could be greater than unity! Such a question would be improper, because,
{\em we never used either the specific form of External SM} ($g_{RR} = -
[1-R_{g}/R]^{-1}, g_{TT} = [1-R_{g}/R])$ or {\em any other specific form for}
$A$ and $B$. All we can try to do to ward off such likely prejudices and
lingering suspicions, is to  remind again that by
definition worldlines of material particles are timelike and the
singularity theorems deal only with such worldline. So one can take a
timelike radial worldline $ds^2 >1$ in the background of a
general dynamic coordinate system:
\begin{equation}
ds^2 = A^2 (dx_0)^2 - B^2 dr^2 >0
\end{equation}
or,
\begin{equation}
A^2 (dx^0)^2 \left[ 1 -  {B^2 dr^2 \over A^2 (dx^0)^2}\right] \ge 0
\end{equation}
In the interior of the fluid (or anywhere if a proper coordinate is used,
or, in the present case, if External SM metric coefficients are not used),
$A^2 >1$, and therefore
\begin{equation}
 \left[ 1 -  {B^2 dr^2 \over A^2 (dx^0)^2}\right] \ge 0
\end{equation}
Or,
\begin{equation}
1 -v^2 \ge 0; \qquad v^2 \le 1
\end{equation}
And in any case in Sec. IX.B. , we have shown that even if there would be
any coordinate singularity, our central result remains intact because the
determinant of the metric $g$ is always negative.
\section{Revisiting Finite Mass Black Holes!}
Even if, in view of the present work, it appears that  finite
mass BHs can not be generated by GTR goverened gravitational collapse,
some readers may think that
 we may not absolutely rule out  the existence of such objects in the
physical universe and continue studying a plethora of BH
related interesting physics either in the context of gravity or say
superstring theories.

If a BH is allowed in physics, of course, the actual worldline of a
test particle or anything inside the event  horizon must be timelike. And
{ \em this is believed to be ensured by shifting to appropriate coordinate system}.
Let us try to verify whether
it is indeed the case with respect to Lemaitre coordinates.

Since the Lemaitre coordinates are believed to be
valid only for $R \le R_{gb}$,  for $R >R_{gb}$ region,  the External SM
is perfectly valid. If really so, the expression $v_{ex} = (R_{gb}/R)^{1/2}$
should be valid in a region infinetisimally close to $R_b=R_{gb}$, so that, the
value of $v_{ex}$ was allowed to be arbitrarily close to $c$ without being
equal to $c$. If so, what would be the value of $v$ in the Lemaitre
coordinate at $R_b=R_{gb}$? Even though the old expression for $v_{ex}$ is no
longer valid, the physical velocity
 experienced by a comoving observer, a velocity which would occur in his
local Lorentz transformation equations, must {\em monotonically increase
in a spherical geometry
in the presence of the acceleration
induced by the central singularity}. Therefore if it was already
infinitesimally close to $c$, how can we demand that $v$ would remain so
and would not exceed $c$ as the observer
traverses from $R> R_{gb}$, and then  to $R\rightarrow 0 $, if $R_g$ is
indeed finite? Let us try to find this in an  explicit manner.

And while doing so, again {\em we would avoid explicitly bringing in the
concept of any physical 3-velocity} $v$ because many readers or experts
might be confused about its definition.

\subsection{Lemaitre Coordinate}
Lemaitre coordinate system is supposed to be the actual
COF for the region
$R \le R_{gb}$:
\begin{equation}
ds^2 = d\tau^2 + g_{rr} dr^2 -R(r,\tau)^2 (d\theta^2 +\sin^2 \theta d\phi^2)
\end{equation}
where
\begin{equation}
-g_{rr} = \left[{3\over 2 R_{gb}} (r - \tau)\right]^{-2/3}
\end{equation}
and
\begin{equation}
R = \left[{3\over 2} (r -\tau)\right]^{2/3} R_{gb}^{1/3}
\end{equation}
Here,
\begin{equation}
 r-\tau \rightarrow
(2/3) R_{gb};\qquad  as~ R \rightarrow R_{gb}
\end{equation}
and for the central singularity, we have
\begin{equation}
r-\tau \rightarrow o;\qquad as~  R \rightarrow 0
\end{equation}
Evidently, this new metric is regular
everywhere except at $R=0$ where the physical singularity is present
provided $R_{gb} >0$. The
total spacetime is now analyzed by two a piece of coordinate system, (1)
External SM for $R > R_{gb}$ and (2) Lemaitre Coordinate for $ R\le R_{gb}$.
And, in this hybrid coordinate the entire spacetime is (except $R=0$
point) is believed to be well behaved with {\em all geodesics as
timelike}, as they must be by  the postulations of GTR.

Let us reexpress  the Lemaitre metric, for a radial worldline as
\begin{equation}
ds^2 = d\tau^2\left[ 1 + {g_{rr} dr^2\over d\tau^2}\right]
\end{equation}
Or,
\begin{equation}
ds^2 = d\tau^2\left\{1 - \left[{3\over 2 R_{gb}} (r - \tau)\right]^{-2/3}
 {dr^2\over d\tau^2}\right\}
\end{equation}
Now by differentiating Eq.(11.4), we find
\begin{equation}
{dr\over d\tau} \rightarrow 1; ~as ~ R\rightarrow R_{gb}
\end{equation}
so that
\begin{equation}
ds^2 = d\tau^2 \left\{[1 - \left({3\over 2 R_{gb}} (r -
\tau)\right]^{-2/3}\right\}; \qquad R\rightarrow R_{gb}
\end{equation}
Now by insering Eq. (11.4) in the above Eq. we find
\begin{equation}
ds^2 = d\tau^2 ( 1 - 1) = 0; ~as ~ R\rightarrow R_{gb}
\end{equation}
This means that the {\bf metric has become null} even in the correct coordinate system

Further by differentiating Eq. (11.5), we find that for the central singularity, $R\rightarrow 0$ too
\begin{equation}
{dr\over d\tau} \rightarrow 1;\qquad as ~ R\rightarrow 0
\end{equation}
And then using Eq. (11.5) in Eq.(11.9), we find that
\begin{equation}
ds^2 = d\tau^2 ( 1 - \infty) = -\infty;\qquad as ~ R\rightarrow 0
\end{equation}
This means that the {\bf metric has become spacelike} even in the correct coordinate system

Note {\em we did not introduce any concept of} $v$ in this simple derivation which
directly shows that the concept of a BH is not allowed by GTR. The results
$ds^2=0$ and $ds^2 =- \infty$ may however be physically explained by
stating that $v=1$ in the first case and $v=\infty$ in the seccond case.
This is exactly what happens for a BH in Newtonian physics and it shows that
the concept of BH is essentially a Newtonian one. For, GTR,
such anomalies can be techinically eliminated only for the case of $R_{gb}
=0$ or if the mass of the BH is zero.

 It is a great irony that
nobody ever tried to verify that even after the desired coordinate
transformation, the same anomalies which plagued the External SM $v^2_{ex}
\rightarrow  R_{gb}/R \rightarrow \infty$ are very much present. And this unacceptable
features can be removed if and only if $R_{gb} \equiv 0$ or
if the {\bf socalled $T$
region is banished from physics}. Now let us examine the case of the
Kruskal coordinate.

\subsection{ Kruskal Coordinate}
In terms of a coordinate system, $(r_*, t_*)$, possessing rather unusual properties:
\begin{equation}
r_*^2 - t_*^2 = K^2 \left( {R\over R_{gb}} -1\right) \exp \left({R\over R_{gb}}\right)
\end{equation}
and,
\begin{equation}
{2 r_* t_* \over r_*^2 + t_*^2} \equiv \tanh \left ({T\over R_{gb}}\right)
\end{equation}
where $K$ is an arbitrary constant. The metric is
\begin{equation}
ds^2={4 R_{gb} \over R K^2} \exp \left(- {R\over R_{gb}}\right) (dt_*^2 - dr_*^2) -
R(r_*, t_*)^2 (d\theta^2 + d\phi^2 \sin^2 \theta)
\end{equation}
So, here, we have,
\begin{equation}
-g_{r_* r_*} =g_{t_* t_*} = {4 R_{gb} \over R K^2} \exp \left(- {R\over R_{gb}}\right)
\end{equation}
Again we rewrite the radial worldline in a Kruskal metric as
\begin{equation}
ds^2={4 R_{gb} \over R K^2} \exp \left(- {R\over R_{gb}}\right) (dt_*^2 - dr_*^2)
={4 R_{gb} \over R K^2} \exp \left(- {R\over R_{gb}}\right) dt_*^2\left( 1 -
{dr_*^2\over dt_*^2}\right)
\end{equation}
Here the event horizon corresponds to
\begin{equation}
r_*^2 -t_*^2 \rightarrow 0; ~as~ R\rightarrow R_{gb}
\end{equation}
and the central singularity corresponds to
\begin{equation}
r_*^2 \rightarrow t_*^2  - K^2; ~ as ~ R\rightarrow 0
\end{equation}
By differentiating the above two equations, we obtain, in either case
\begin{equation}
{dr_*\over dt_*} \rightarrow {t_*\over r_*} ;\qquad as, ~ R\rightarrow R_{gb}
~or~ R\rightarrow 0
\end{equation}
Then for both these regions, the metric, Eq.(11.17), may be rewritten as
\begin{equation}
ds^2
={4 R_{gb} \over R K^2} \exp \left(- {R\over R_{gb}}\right)
 \left({r_*^2 -t_*^2\over r_*^2}\right) dt_*^2
\end{equation}
Therefore, as the horizon is approached, by using Eq.(11.18) into
 Eq.(11.20) we find that,
\begin{equation}
ds^2
={4  \over  K^2} \exp (- 1) dt_*^2 (0) = 0
\end{equation}
Further by using Eq.(11.19) in Eq.(11.20), we find that
 as the central singularity is approached ($R\rightarrow 0$)
\begin{equation}
ds^2
={4 R_{gb} \over R K^2} \exp \left(- {R\over R_{gb}}\right)
\left({-K^2\over r_*^2}\right) dt_*^2
\end{equation}
Thus, clearly we find that not onlt has the metric
blown up at $R\rightarrow 0$, but  {\bf it has become spacelike} too:
\begin{equation}
ds^2 = - (\infty)   {dt_*^2\over r_*^2} = - \infty
\end{equation}
 So, as before, the {\bf radial geodesic
becomes null and then spacelike even in the Kruskal coordinate}. And all
these difficulties can be resolved if and only if $R_{gb} =0$, i.e, when we
realize that there is no event horizon all.

 If so, all the curvature components and the
associated scalars can be seen to blow up at $R = R_{gb} =0$, the true singularity.
This understanding would free physics of the riddle of the {\em true nature} of the Schwarzschild Singularity.
It is unbelievable that rather than thinking in this way, we have all
along allowed us to be swayed by the apparent regularity of the Lemaitre
or Kruskal metrics.

 Again this points  to the fact that the goal of removing the
singularities were really not achieved and all that what was actually
achieved by such efforts were of purely cosmetic nature. The real
difficulty lay at a much more fundamental level, and in the consequent
 incorrect premises of the problem which presumed the existence
of a finite mass BH. It is unfortunate that rather than delving into the
real reason behind the occurrence of the Schwarzschild Singularity many
authors have gone even one step further, and
have seriously pursued the notion that there is a $T$ region where
the preexisting space coordinate $R \rightarrow {\tilde T}$, a weird time
coordinate and the (External) Schwarzschild time coordinate $T \rightarrow
{\tilde R}$, some weird space coordinate. This was pursued in view of
the fact that in the External SM,
$g_{RR}$  and $g_{TT}$ would exchange their respective signs if the Event
Horizon would be crossed.  However, this was unjustified because,
{\bf even in the Lemaitre and Kruskal coordinates the angular part of the
metric explicitly involved the same Schwarzschild circumference coordinate}
$R$ and not any weird spatial coordinate ${\tilde R} \sim T$.

It is satisfying to recall that atleast one physicist has expressed his
reservation about the reality of the region inside the event horizon
without any ambiguity [33]:

``so that in this region $R$ is timelike and $T$ is spacelike. However,
this is an impossible situation, for we have seen that $R$ is defined
in terms of the circumference of a circle so that $R$ is spacelike, and we
are therefore faced with a contradiction. We must conclude that the
portion of space corresponding to $R <2M$ is non-physical. This is a situation
which a coordinate transformation even one which removes a singularity can
not change. What it means is that the surface $R=2M$ represents the
boundary of physical space and should be regarded as an impenetrable
barrier for particles and light rays.''

And when we realize that trapped surfaces or event horizons can not occur
in Nature if GTR is a correct physical theory, we would be instantly able
to resolve the debate between physicsts that when the existence of BHs and
event horizons imply loss of information the from observable universe, in
violation of the premises of Quantum mecahanics, how can one have a
successful theory of Quantum Gravity which incorporates GTR at the
classical level.

\section{Conclusion}
The important work of Oppenheimer and Snyder [26] which gave the
(incorrect) impression of
formation of a BH of finite mass $M_b$ and event horizon $R_{gb}$ in a
comoving proper time $\tau_{gb} \propto R_{gb}^{-1/2} \propto
M_b^{-1/2}$  was technically correct
except for the fact that it tacitly assumed  $M_b =M_f =M_i \approx M_0$.
 Actually, the equation (36) of their paper (Eq. [8.9] in the present
paper) demands that
in order that, at the boundary of the star,
\begin{equation}
T \sim \ln {y^{1/2} +1
\over y^{1/2} -1} \ln {(R_b/R_{gb})^{1/2} +1
\over (R_b/R_{gb})^{1/2} -1}
\end{equation}
 remains {\bf definable}, one
must have $R_{gb} < R_b$. And then the central singulaity $R_b \rightarrow
0$ could be reached only if $R_{gb}=0$, if the horizon coincides with the
central singularity. And then there would be no region (T-region)
interior to the horizon.  Accordingly, the value of $\tau_{gb} = \infty$
along with $T=\infty$. In fact this result {\em follows in a trivial fashion}
from Eq. (32) of their paper (our Eq.[8.10])
\begin{equation}
y\equiv {1\over 2} [(r/r_b)^2 -1] + {r_b \over r} {R\over R_{gb}}
\end{equation}
where the parameter $y$
which {\bf must be positive}. But if $R_{gb} \neq 0$,
 as $R\rightarrow 0$, it is trivial to see that $y$
 {\bf actually becomes negative} for $r <r_b$. This shows that actually the horizon or any trapped
surface in never allowed by the OS solution. And this resolves the following puzzle. The
physical feeling of ``time'' for two different observers is of course not
 absolute in either STR or GTR. But it does not mean that, for non-quantum
classical physics, we can have a ``Schrodinger's cat paradox'' like
scenario. Here two observers may differ on the ``size'' ``weight'' and even
the ``age'' of the cat. But, if one of them finds the cat to be ``dead''
the other observer, probably, can not find it to be alive and
kicking for ever.

  More importantly, the
formulation of the problem of homogeneous dust collapse is faulty because
it corresponds to $N=0$. In an independent and general manner we reached
the same conclusion about the problem of collapse of
spherical inhomogeneous dust too.

Moving away from the idealized dust solutions, we found that for the
continued collapse of any perfect fluid possessing arbitrary EOS and
radiation transport properties, a proper amalgamation of the inherent
global constraints arising because of the dependence of spatial curvature
like parameter ($\Gamma$) on the global mass-energy content, $M$,
 directly shows that {\em no trapped surface is
allowed by GTR}. This result becomes independent of the details of the
radiation transport properties because the integration of the (0,0)-
component of the Einstein equation, yields the definition of $M$ by
absorbing all  quantities like $\rho$,  $q$, and $H$, in whichever fashion they
may be present. Then it follows that if there is a continued collapse, on
whatever time scale it might be, the final gravitational mass of the
configuration necessarily become zero.  This $M_f=0$ state must not be
confused as a vacuum state, on the other hand, the baryons and leptons are
crushed to the singularity with an infinite negative gravitational energy $E_g
\rightarrow -\infty$. On the other hand, the
positve internal energy is also infinite $E_{in} \rightarrow \infty$, but
seperated by the energy gap $\rightarrow M_0$.
However, in the present paper, we did not investigate whether this state
corresponds to $2GM_b/R <1$ or $2GM/R =1$. In another work [3],
 we find that, it is the latter limit which should be appropriate, i.e.,
the system keeps on radiating and tends to  attain the
state of a zero mass  BH characterized by zero energy and entropy, the
ultimate ground state of classical physics. Neither did we try to find
here the proper time required to attain this absolute classical singular
ground state though we found that for the fictitious dust solutions $\tau =\infty$.
This question has, however, been explored elsewhere [3] to find that, for a
real fluid too, $\tau =\infty$. This means that there is no incompleteness
in the radial worldlines of the collapsing fluid particles inspite of $R$
having a finite range ($r_b$). Such a Non-Newtonian behavior is
understandable in GTR because, it was found [3] that although
$M$ keeps on decreasing, the curvature components $\sim GM/R^3 \sim R^{-1}$ tend to
blow up. As a result the 3-space gets stretched and stretched by the strong
grip of gravity, or in other words, the proper distances eventually tend to blow up too.

We also explicitly showed that, if one assumes the existence of a
Schwarzschild BH of finite mass $M_b$,  the actual worldlines of a ``test
particle'', taken with reference to Lemaitre coordinate or Kruskal coordinate,
  would really become  spacelike with the physical speed $
v\rightarrow \infty$ as $R\rightarrow 0$, in complete violation of STR.
This simple fact independently asserts that
{\bf there is no Event Horizon, no  Schwarzschild Singularity, no T-region},
 and the only singularity that might have been present is the central singularity,
and {\bf whose mass must be zero}. And technically, one might view this
central singularity as the Schwarzschild Singularity associated with a
zero mass BH. Even then the existence of such a zero mass BH could be
realized only if the collapse process could be complete in a finite proper
time; but it actually takes infinite time : Nature abhors not only naked
singularities but all singularities; and we find that only GTR may be having the
mechanism of removing such singularities even at a classical level. And
this happens because of the marriage between the physics and space(time)
geometry. If somehow, one would try to
build up a super concentrated energy density
near a ``point'', the space would get dynamically stretched by the gravity associated with
the concentrated energy density  and a singularity is avoided.

Consequently, all the associated theoretical
confusions like (i) whether the physically defined circumference coordinate $R$
can, suddenly become a time-like coordinate, (ii) whether there could be White
Holes freely spewing out matter and energy in the observable universe,
(iii) whether there could be macroscopic Worm Holes providing short cut to distant
regimes of spacetime, and (iv) whether information can really be lost from
the observable universe in violation of the quantum mechanics, which have
plagued GTR in the present century, would be resolved, if the present work is
correct.

Finally, we appreciate the physical intuition of Einstein [1] and
Landau [29] in
not being able to accept the reality of Schrawzschild Singularity
or any singularity in GTR. We again recall that  Rosen [33], in an
unambiguous manner noted the impossible and unphysical nature of the T-region.
We have, in this paper, resolved all such paradoxes by showing that not only
the $R <2M$ region unphysical, it does not exist or is not ever created.

Although, it might appear that  astrophysics would be poorer in the
absence of the mystique of BH, actually, it may be possible to envisage
new varieties of  stable or quasi- stable ultracompact compact objects of
stellar mass or
dynamically contracting super massive stars responsible for new gamut of
astrophysical phenomenon. To appreciate this statement, it is necessary to
{\em feel} (everybody knows it) that, $M_{OV}$ {\bf refers to the maximum
allowed mass} and not the {\bf minimum possible mass}. Oppenhemier and
Volkoff equation or any equation does not really yield any   lower limit
on the {\em gravitational mass} of a NS (or the compact object).
And the lowest limit is  $M_f=0$.
The usual lower limit of a NS
that is discussed in the literature actually refers to the baryonic mass
If one instinctively invokes the incorrect form of $M_{OV}$ involving
$M_0$ (Eq. 1.2), one would not be able to go beyond the standard idea that a
star having a main sequence mass, say, $M_0 > 10 M_\odot$ can not end up
as a NS. On the other hand, when we use the correct form of $M_{OV}$
involving the final gravitational mass $M_f < M_0$, in principle, stars
with much larger initial main sequence mass $M_0 > 10 M_\odot$ may
collapse to become a NS of mass either equal to or smaller than $M_{OV}$ by any  amount.
However, such a NS may have much lagrer baryon density than a canonical NS.
Yet, there is one constraint imposed by  GTR on such ultracompact objects [48], which
demands that if the compact object is assumed to be {\em cold} and in
hydrostatic equilibrium, the surface redshift
\begin{equation}
z_s = \left({1\over \Gamma} -1\right) <2
\end{equation}
It does not mean that there can not be any compact object beyond this
limit, i.e., $z_s \ge 2$. It only means that such high $z_s>2$ objects must be ``hot''
and probably dynamically contracting (remember the time to collaes to a
singularity is $\infty$).

Also recall here that if rotation is taken into considerations, the value
of $M_{OV}$ could be significantly raised. But to fully appreciate the
question of likely existence of stellar mass BH candidates of masses as
high as $\sim 10 M_\odot$, we must keep an open mind with regard to our
present day understanding of QCD. Even with reference to present state of
knowledge of QCD, there could be compact objects with exotic EOS, where
the masses could be $\sim 10 M_\odot$ or even higher [23]. These stars are
called Q-stars (not the usual quark stars), and they could be much more compact than
a canonical NS; for instance,
a stable  non rotating Q-star of mass 12$ M_\odot$ might have a radius of
$\sim 52$ Km [23]. This may be compared with the value of $R_{gb} \approx
36$ Km of a supposed BH of same mass.

In general, it is believed that, at sufficient high temperature, quark
confinement may melt away. And the energy gained from the pairing of quarks
and antiquarks of all colors which drive the chiral symmetry breaking may
be overcome by the entropic advantage in letting the particles be free. At
a very high $T$, therefore, asymptotically, free quarks, antiquarks and
gluons should be liberated [17] and provide new sources of pressure. There is already
some
 evidences that at a temperature of $\sim 150$ MeV,
there is  a phase transition in hot nuclear matter and new degrees of
freedom are suddenly liberated [17]. It is such processes which may allow
ultracompact objects to be in a stable or dynamic quasi-stable state.

Let us briefly recall the case of the recently discovered unusual
supernova 1998bw [38], whose ejecta is approximately 10 times more
energetic than normal supernova ejecta, finds no explanation in terms of the canonical idea that
the gravitational collapse of stellar mass objects can release a maximum
energy of $\sim 10^{53}$ erg because higher mass cores quietly become a BH
without releasing appreciable energy. Similarly, some of the optically detected
 Cosmological Gamma Bursts like GRB970508 and GRB971214 having an energy
$Q_\gamma > 10^{53}$ erg in the gamma-rays alone may require an
original energy output, in the form of a neutrino burst somewhere in the
region of
$Q_\nu \sim 10^{54-55}$ erg [4] and for which there is no proper explanation in
the present paradigm. In fact the Gamma Ray Burst of 23rd Janurary, 1999
has radiated an amount of $Q_\gamma \approx 2.3 \times 10^{54}$ erg (under condition of
isotropy), and presumably much more in neutrinos. Such energy release is
hardly possible if trapped surfaces really formed at values of
$M_f \approx M_i$. On the other hand such phenomenon might be signaling
the formation of new relativistic ultracompact objects.

\section {Acknowledgement}

 I am thankful to Prof. P.C. Vaidya for carefully
going through an initial version of this work, offering some suggestions, and confirming that
this work is physically and general relativistically correct.


\newpage
\centerline{\bf APPENDIX 1}
\vskip 1cm

This appendix contains photostat of pp. 249-252, and pp. 311 from {\em The
Classical Theory of Fields} by L.D. Landau \& E.M. Lifshitz, 4th Edition
(Pergamon, Oxford 1975).

The first 4 pages 249-252 are from Section 88 titled ``The constant
gravitational field''. It shows that we have indeed used the correct
expression for the locally measured 3-velocity $v= {dl\over d\tau}$, where
$dl$ is the proper distance and $d\tau$ is the proper time. On the other hand, pp.
311 is from Section 102, and at its bottom shows the specific form of
$v^2$ for spherical symmetry. Here the radial variable is indicated by the
subscript ``1'' in place of ``$r$'' used by us.

\newpage
\centerline{\bf APPENDIX 2}
\vskip 1cm

This appendix contains photostat of pp. 94 from {\em Relativistic
Astrophysics}, Vol. 1 by Y.B. Zeldovich \& I.D. Novikov. (Univ. Chicago,
Chicago, 1971).

The top portion of this page mentions about the physical velocity
$v={dx\over d\tau}$, where $dx$ is element of proper radial distance and
$d\tau$ is element of proper time. Note that following Landau \& Lifshitz,
we have used the nomenclature ``$dl$'' for proper distance. It is also
discussed here that it is this $v$ which appears in the local Lorentz
transformation. However, it should be reminded that Zeldovich \& Novikov's
discussion is in the context of the External Schwarzschild Metric.

\newpage
\centerline{\bf APPENDIX 3}
\vskip 1cm

This appendix contains photocopy of pp. 675 from {\em Gravitation} by C.W.
Misner, K.S. Thorne \& J.A. Wheeler (Freeman, San Fansisco, 1973).

The mid-portion of this page shows the expression for the ``ordinary''
velocity, i.e, the physical 3-velocity. For a purely radial motion,
$v_\phi =0$ and $v=v_r$.
as measured by a local observer

\newpage
\centerline{\bf APPENDIX 4}
\vskip 1cm

This appendix contains photocopy of pp.75 from {\em Gravitation Theory and
Gravitational Collapse} by B.K. Harrison, K.S. Thorne, M. Wakano, and J.A.
Wheeler, (Univ. Chicago, Chicago, 1965).

The bottom portion of this page discusses the idea that for arbitrary
equation of state it is possible  to have a final state having
gravitational mass $M=0$ (They used asterisk to denote gravitational mass
to differentiate from baryonic mass).

\newpage
\centerline{\bf APPENDIX 5}
\vskip 1cm

This appendix contains photocopy of pp.297 from {\em Relativistic
Astrophysics}, Vol. 1 by Y.B. Zeldovich \& I.D. Novikov. (Univ. Chicago,
Chicago, 1971).

The authors specifically discuss here the possibility of having an
ultradense baryonic configuration whose ``mass defect'' is equal to the
baryonic mass indicating that the final gravitational mass $M_f =0$. They
also correctly point out that such a scenario might be achieved only by
for dynamic collapse.

\end{document}